\title{On quantum group symmetry and Bethe ansatz
for the \AC\ spin chain with integrable boundary}
\author{Anastasia Doikou\footnote{e-mail: doikou@lapp.in2p3.fr}
\\ {{\footnotesize LAPTH, Annecy--Le--Vieux, BP 110 F-74941, France}}
\\
\\ Paul P. Martin\footnote{e-mail: p.p.martin@city.ac.uk}
\\ {{\footnotesize City University, Mathematics Department,
Northampton Square, London EC1V 0HB, UK}}}
\date{}
\documentclass[12pt]{article}
\usepackage{epic}
\usepackage{eepic}
\usepackage{graphicx}
\usepackage{latexsym}
\usepackage{amssymb}
\usepackage[all]{xy}
\usepackage{citesort}
\providecommand{\noglossaryignore}[1]{}
\newcommand{\globalglossaryentry}[3]{\makebox[1.5in][l]{\tt $\backslash${#1}} 
\makebox[1.1in][l]{{$#2$}} \makebox[2.5in][l]{{#3}}\newline} 
\newcommand{\newcommandabbreviation}[3]{\newcommand{#1}{#2}%
\noglossaryignore{\globalglossaryentry{#3}{#2}{}}}
\newcommand{\renewcommandabbreviation}[3]{\renewcommand{#1}{#2}%
\noglossaryignore{\globalglossaryentry{#3}{#2}{}}}
\newcommand{\newcommandmacro}[4]{\newcommand{#1}{#2}%
\noglossaryignore{\globalglossaryentry{#3}{#2}{#4}}}
\newcommand{\gge}[3]{\noglossaryignore{\globalglossaryentry{#1}{#2}{#3}}}
\newcommand{\myaddress}%
{\parbox{3in}{\footnotesize \begin{center} 
Mathematics Department, City University, \\  
Northampton Square, London EC1V 0HB, UK.\end{center}}}
    
\newcounter{minidef}[section]

\newcounter{minicapt}

\newtheorem{theo}{Theorem}         
\newtheorem{de}{Definition}     \newtheorem{pr}{Proposition} 
\newtheorem{co}{Corollary}[pr]   
\newtheorem{lem}{Lemma} 

\noglossaryignore{GREEK ETC.\newline}
\newcommandabbreviation{\e}{\epsilon}{e}        
\newcommandabbreviation{\lam}{\lambda}{lam}  
\newcommandabbreviation{\la}{\langle}{la}        
\newcommandabbreviation{\ran}{\rangle}{ran}
\newcommandabbreviation{\ha}{\#}{ha}             
\newcommandabbreviation{\rmap}{\rightarrow}{rmap}
\newcommandabbreviation{\aaa}{\alpha}{aaa}        
\newcommandabbreviation{\ab}{\alpha,\beta}{ab}
\newcommandabbreviation{\aab}{a(\ab )}{aab}       
\noglossaryignore{\newline RINGS\newline}
\newcommandabbreviation{\HH}{H \!\!\! I}{HH}              
\newcommandabbreviation{\C}{\mathbb C}{C}
\newcommandabbreviation{\N}{\mathbb N}{N}  
\newcommandabbreviation{\Z}{\mathbb Z}{Z}     
\renewcommandabbreviation{\Re}{\mathbb R}{Re}
\newcommandabbreviation{\R}{{\mathbb R}}{R}
\newcommandabbreviation{\Q}{\mathbb Q }{Q}
\renewcommandabbreviation{\H}{\mathbb H }{H}
\noglossaryignore{\newline SYMMETRIC GROUP\newline}
\def\Sym(#1){\Sigma(#1)}                  
\gge{Sym(-)}{\Sym(-)}{symmetric group on - objects}
\def\Sy(#1){\Sigma_{#1}}                  
\gge{Sy(-)}{\Sy(-)}{symmetric group irreducible -}
\def\sym(#1){\mbox{\LARGE s}(#1)}       
\gge{sym(-)}{\sym(-)}{symmetric group on - objects (variant)}
\def\sy(#1){\mbox{\LARGE s}({#1})}       
\gge{sy(-)}{\sy(-)}{symmetric group irreducible - (variant)}
\newcommandmacro{\cs}{\C \, \sy(n)}{cs}{symmetric group algebra over $\C$}
\noglossaryignore{\newline PARTITIONS/SETS\newline}
\newcommand{\Nset}[1]{\underline{#1}}
\gge{Nset\{-\}}{\Nset{-}}{set of natural numbers to -}
\def\nset(#1){ \{ #1 \}_{ \underline{n} }}
\gge{nset(-)}{\nset(-)}{a set $-\times\Nset$}
\def\ul(#1){_{\underline{#1}}}            
\gge{ul(-)}{{}\ul(-)}{subscript underline -}
\def\Ee(#1){{\bf E}_{#1}}                 
\gge{Ee(-)}{\Ee(-)}{set of equivalence relations on set -}
\def\Eee(#1){{\bf E}_{\{ #1 \}_{\underline{n}}}}  
\gge{Eee(-)}{\Eee(-)}{ditto for nset}
\def\Een(#1,#2){{\bf E}_{\{ #1 \}_{\underline{#2}}}}  
\def\Ssn(#1,#2){{\bf S}_{\{ #1 \}_{\underline{#2}}}}  
\def\Ss(#1){{\bf S}_{#1}}                 
\def\Sss(#1){{\bf S}_{\{ #1 \}_{\underline{n}}}}  
\def\bbc(#1){((\beta_1)(\beta_2)...(\beta_{#1}))}     
\newcommandmacro{\Ln}{{\Gamma}^{n}}{Ln}{large index set}
\newcommandmacro{\LnQ}{{\Gamma}^{n}_Q}{LnQ}{index set}
\newcommandmacro{\Zz}{\zeta}{Zz}{`shape' function}
        
\noglossaryignore{\newline PARTITION ALGEBRA\newline}
\def\ka(#1){\kappa_{#1}}                  
\def\Sm(#1){\Sigma_{#1}}                  
\newcommandmacro{\com}{\bullet}{com}{bullet composition}
\newcommandmacro{\enm}{\; e^n(\! m\! ) \;}{enm}{product of idempotents}
\def\Ai(#1){ A^{ #1 \cdot } }             
\def\Aij(#1,#2){ A^{ #1  #2 } }           
\newcommandmacro{\One}{\mbox{\bf $1 \!\!\! 1$}}{One}{algebra unit 1}

\newcommandmacro{\Bp}{B_p}{Bp}{partition basis}
\def\Bb(#1){B_p[#1]}                      
\def\Pp(#1){P_n[#1]}                      
\def\Ps(#1){P_n[#1] \! /}                 
\newcommandmacro{\Ph}{\hat{P}}{Ph}{P hat  algebra}
\def\Is(#1){\sim^{#1}}                    
\noglossaryignore{\newline MODULES\newline}
\def\Wm(#1){{\cal S}_{#1}}                
\gge{Wm(-)}{\Wm(-)}{Weyl module with index -}
\def\wm(#1,#2){{}_{#1}{\cal S}_{#2}}      
\gge{wm(-1,-)}{\wm(-1,-)}{Weyl module with index -}
\def\Ind(#1,#2,#3){\mbox{Ind}_{#1}^{#2}#3}
\gge{Ind(-1,-2,-)}{\Ind(-1,-2,-)}{induction}
\def\Res(#1,#2,#3){\mbox{Res}_{#1}^{#2}#3}
\gge{Res(-1,-2,-)}{\Res(-1,-2,-)}{restriction}
\newcommandabbreviation{\weyl}{standard}{weyl}
\newcommandabbreviation{\mod}{\mbox{mod}}{mod}
\newcommandabbreviation{\head}{\mbox{head }}{head}
\newcommandabbreviation{\Weyl}{Weyl}{Weyl}
\def\SS(#1){{\cal S}_{#1}}                
\gge{SS(-)}{\SS(-)}{Specht/Weyl module index -}
\def\LL(#1){{\cal L}_{#1}}                
\gge{LL(-)}{\LL(-)}{simple module index -}
\noglossaryignore{\newline FUNCTORS/MAPS\newline}
\newcommandmacro{\Gg}{{\cal G}}{Gg}{G Functor}
\newcommandmacro{\Fg}{{\cal F}}{Fg}{F Functor}
\newcommandmacro{\ra}{\rightarrow}{ra}{}
\def\ses(#1,#2,#3){0\ra #1 \ra #2 \ra #3 \ra 0}  
\gge{ses(1,2,-)}{\ses(1,2,-)}{\hspace{.5in} short exact sequence}
\def\starr(#1){ \stackrel{ #1 }{\longrightarrow} }
\gge{starr(-)}{\starr(-)}{}
\newcommandmacro{\doublerightarrow}{\; -\!\!\! -\!\!\!\!\!\! \gg \;}
{doublerightarrow}{}
\noglossaryignore{\newline PARTITION ALGEBRA MAPS\newline}
\newcommandmacro{\smap}{s}{smap}{`inclusion' map}
\newcommandmacro{\tmap}{t}{tmap}{$ P_n -> S_n$}
\newcommandmacro{\pmap}{\psi}{pmap}{$ S_n -> P_n $}
\noglossaryignore{\newline MISC.\newline}
\def\Amap(#1){{\cal A}_{#1}}              
\gge{Amap(-)}{\Amap(-)}{}
\def\Rr(#1){R_{#1}}                       
\gge{Rr(-)}{\Rr(-)}{restriction of E}
\def\Cr(#1){C_{#1}}                       
\gge{Cr(-)}{\Cr(-)}{restriction of E to N}
\newcommandmacro{\Tm}{{\cal T}}{Tm}{Transfer Matrix}
\def\On(#1){{\cal I}_{#1}}
\gge{On(-)}{\On(-)}{}
\newcommandmacro{\UU}{\underline{\sqcup}}{UU}{}  
\newcommandmacro{\UUU}{\sqcup}{UUU}{}  
\newcommandmacro{\Vq}{V_Q^{\otimes n}}{Vq}{Potts config. space}
\def\bs(#1,#2){\mbox{{\Large $\ast$}}^{#1}_{#2}} 
\gge{bs(-,-)}{\bs(-,-)}{general plumbing multiplier}
\newcommand{\ignore}[1]{}
\newcommand{\hignore}[1]{}
\gge{ignore\{-\}}{\hignore{-}}{ignore argument!}
\def\choo(#1,#2){ \left( \begin{array}{c} #1 \\ #2 \end{array} \right) }
\gge{choo(-1,-)}{\choo(-1,-)}{choose}
\newcommand{\Qed}{$\Box$}
\gge{Qed}{\mbox{\Qed}}{QED}
\def\staq(#1){\stackrel{#1}{=}}           
\gge{staq(-)}{\staq(-)}{}
\def\stam(#1){\stackrel{#1}{\rightarrow}} 
\gge{stam(-)}{\stam(-)}{}
\def\mat{ \left( \begin{array} }    
\def\tam{ \end{array}  \right) }
\gge{mat/tam}{...}{matrix delimiters}
\newcommand{\beq}{\begin{equation} }
\def\eql(#1){ \begin{equation} \label{#1} 
}
\newcommand{\eq}{\end{equation} }
\def\eqal(#1){\begin{eqnarray} \label{#1} }
\def\eqa{\end{eqnarray} }
\def\lab(#1){\label{#1}
}
\def\prl(#1){ \begin{pr} \label{#1} 
}
\def\del(#1){ \begin{de} \label{#1} 
}
   
\gge{smeq\{-\}}{...}{small equation}
  
\gge{fneq\{-\}}{...}{very small equation}
\noglossaryignore{\newline HECKE/BLOB\newline}
\newcommandmacro{\Hnq}{H_n(q)}{Hnq}{ * freestanding symbol}
\newcommandmacro{\Hn}{H_n}{Hn}{      *-mod etc.}
\newcommandmacro{\A}{{\cal A}}{A}{}
\newcommandmacro{\Cwts}{C}{Cwts}{}
\newcommandmacro{\CA}{{\cal A}}{CA}{}

\newcommandmacro{\calA}{{\cal A}}{calA}{}
\newcommandmacro{\modi}{\mbox{Mod} }{modi}{was mod not modi!}
\newcommandmacro{\Wgen}{{\Bbb S}}{Wgen}{}
\def\ol(#1){\overline{#1}}
\newcommandmacro{\st}{\mbox{St}}{st}{}
  
\def\CMult(#1,#2){(#1:#2)}
\def\CM(#1,#2){( #1 : #2 )}
\def\FMult#1,#2{(#1:#2)}
\def\CF#1,#2{(#1:#2)}

\newcommandmacro{\Top}{\mbox{Top}}{Top}{}
\newcommandmacro{\Soc}{\mbox{Soc}}{Soc}{}
\newcommandmacro{\Head}{\mbox{Head}}{Head}{}
\newcommandmacro{\Filt}{{\cal F}}{Filt}{}
\newcommandmacro{\Mod}{\mbox{mod}}{Mod}{}
\newcommandmacro{\Resi}{\mbox{Res }}{Resi}{was without i!}
\newcommandmacro{\Indi}{\mbox{Ind }}{Indi}{was without i!}
  
\def\RR(#1,#2){R^{#1}_{#2}}  
\def\TT(#1,#2){T^{#1}_{#2}}

\def\Chi{\chi}

\newcommandmacro{\Ann}{\mbox{Ann}}{Ann}{}
\newcommandmacro{\Cen}{\mbox{Cen}}{Cen}{}
\newcommandmacro{\End}{\mbox{End}}{End}{}
\newcommandabbreviation{\semisimple}{semisimple}{semisimple}
\newcommandabbreviation{\Bratteli}{Bratteli}{Bratteli}
\newcommandabbreviation{\JBC}{Jones Basic Construction}{JBC}
\newcommandabbreviation{\pa}{partition algebra}{pa}
\newcommandabbreviation{\TM}{transfer matrix}{TM}
\newcommandabbreviation{\PM}{Potts model}{PM}
\newcommandabbreviation{\QSC}{quantum spin chain}{QSC}
\newcommandabbreviation{\Hamiltonian}{Hamiltonian}{Hamiltonian}
\newcommandabbreviation{\YS}{Young symmetrizer}{YS}

\newcommand{\be}{\begin{eqnarray}}
\newcommand{\eeq}{\end{eqnarray}}
\newcommand{\non}{\nonumber}

\newcommand{\tr}{\mathop{\rm tr}\nolimits}

\newcommand{\YB}{Yang--Baxter}

\newcommand{\yy}{m}

\newcommand{\abU}{{\mathcal U}}
\newcommand{\abe}{{\bf e}}
\newcommand{\Uqsl}[1]{U_q(sl(#1))}
\newcommand{\AUqsl}[1]{U_q(\widehat {sl(#1)})}
\newcommand{\Id}{{\mathbb I}}                   
\newcommand{\Perm}[1]{{\mathcal P_{#1}}}      
\newcommand{\KI}[1]{K^{(#1)}}                   
\newcommand{\rrep}[1]{{\cal R}_{#1}}         
\newcommand{\ij}{i \! + \! 1}
\newcommand{\gemini}{gemini}
\newcommand{\AC}{asymmetric twin}

\begin{document} \maketitle
\newcommand{\ygnore}[1]{}
\newcommand{\ignoreifnotdraft}[1]{#1}
\vskip 2.5cm
\begin{abstract}
Motivated by a study of the crossing symmetry of the  
\AC\ or 
`\gemini'
representation of the  
affine Hecke algebra   
we give a 
construction for {\em crossing tensor space representations} of 
ordinary Hecke algebras. 
These representations   
build solutions to the Yang--Baxter equation 
satisfying the crossing condition 
(that is, integrable quantum spin chains). 
We show that {\em every} crossing representation 
of the Temperley--Lieb algebra appears in this construction, 
and in particular that this construction builds {\em new} representations. 
We extend these to new representations of the blob algebra, which build
new solutions to the Boundary Yang--Baxter equation
(i.e. open spin chains with integrable boundary conditions). 

We prove that the open spin chain Hamiltonian derived from Sklyanin's
commuting transfer matrix using such a solution 
can always be expressed as the representation of
an element of the blob algebra, and determine this element. 
We determine the representation theory (irreducible content) of the
new representations and hence show that all such Hamiltonians have 
the same spectrum up to multiplicity, for any given 
value of the algebraic boundary parameter. 
(A corollary is that our models have the same spectrum as the
open XXZ chain with nondiagonal boundary --- despite differing from 
this model in having reference states.)
Using this multiplicity data, and other ideas, 
we investigate the underlying quantum group symmetry of the new Hamiltonians. 
We derive the form of the spectrum and the Bethe ansatz equations.

\end{abstract}

   \vfill
   \rightline{LAPTH-1089/05}
   \baselineskip=16pt

\newpage
\ignoreifnotdraft{ \newpage } 
\newcommand{\nm}{n}
\newcommand{\rs}{{\hat r}}
\section{Introduction}
   \newcommand{\AR}{A}
   \newcommand{\paula}{\sigma}
   \newcommand{\UUUU}{{\mathcal U}}
The integrability of 
quantum field and lattice theories in two dimensions is closely tied
to the factorization of multi-particle scattering 
\newcommand{\FT}{FaddeevTakhtajan84,FaddeevTakhtajan81}%
\newcommand{\baxter}{Baxter72,Baxter73,Baxter82}%
\cite{McGuire64,FaddeevTakhtajan84,FaddeevTakhtajan81,Zamolodchikov79}. 
Let $R_{ij}(\lambda)$ describe the
scattering amplitude for particles $i,j$ with incidence angle $\lambda$. 
The \YB\ equation 
\cite{McGuire64,Yang67,Baxter72,Baxter73,Baxter82,Korepin80,Korepin93} 
\begin{equation} 
 R_{12}(\lambda_{1}-\lambda_{2})\ R_{13}(\lambda_{1})\ R_{23}(\lambda_{2}) 
 = R_{23}(\lambda_{2})\ R_{13}(\lambda_{1})\
 R_{12}(\lambda_{1}-\lambda_{2}) 
\label{YBE} 
\end{equation} 
provides a key set of
constraints on possible forms of $R_{ij}(\lambda)$ consistent with
factorization. 
When non-trivial
boundaries are present
(e.g. in field theories on a half line), 
$R_{ij}(\lambda)$ must also satisfy  
the boundary \YB\ (reflection)
equation \cite{Cherednik84b,Sklyanin88}:
\begin{equation}
R_{12}(\lambda_{1} -\lambda_{2})\ K_{1}(\lambda_{1})\
R_{21}(\lambda_{1}+\lambda_{2})\ K_{2}(\lambda_{2})
=K_{2}(\lambda_{2})\ R_{12}(\lambda_{1}+\lambda_{2})\
K_{1}(\lambda_{1})\  R_{21}(\lambda_{1} -\lambda_{2}). \label{RE}
\end{equation}
Here the $K$-matrix is the boundary  
scattering matrix of the theory. 

The physical importance of integrable systems with boundary has driven
sustained interest in the study of solutions to these equations
(some key references are
\cite{FringKoberle94a,FringKoberle94b,GhoshalZamolodchikov94a,
deVegaGonzalezRuiz93,%
BatchelorFridkin96,AhnYou97,BehrendPearce96,Gandenberger99,%
DeliusMackay01,DeliusMackay03,DeliusNepomechie02,%
MalaraLimasantos04,BaseilhacKoizumi03}). 
In \cite{DoikouMartin03} the structural similarity between 
these equations 
and the cylinder braid group relations is exploited to derive
solutions systematically. 
In particular, we considered the quotient of
the cylinder braid group algebra called the blob algebra, here denoted
$b_n(q,m)$.
This is a two-parameter extension to the Temperley-Lieb algebra $T_n(q)$. 
The Temperley-Lieb algebra provides a universal approach to 
the \YB\ equation in the sense that 
any tensor space representation 
$\rho: T_n(q) \to \End(V^{\otimes n})$ 
gives a solution to (\ref{YBE}) via a standard construction (see
later), 
and furthermore the Hamiltonian spectrum of each model constructed in this way is
the same (up to multiplicities which may be computed in representation
theory). 
Suppose $\rho$ extends to a representation of $b_n$. 
Then an analogous construction provides a $K$-matrix solving
the reflection equation (\ref{RE}) \cite{LevyMartin94,DoikouMartin03},
and there is an analogous equivalence among corresponding models. 

The simple extension of the ordinary tensor space representation of $T_n$
to $b_n$ which is described in \cite[\S4]{MartinSaleur94a} gives 
the open spin-1/2 XXZ chain with nondiagonal boundary
conditions (see also \cite{deVegaGonzalezRuiz93}). 
This is perhaps the obvious model to start with when studying boundary
conditions, but it has some significant limitations (see later). 
 In \cite{DoikouMartin03} we introduced a model based on an
asymmetrically cabled spin--chain--like representation $\Theta$
of $T_n$, and a particular extension of $\Theta$ to $b_n$. 
The representation is well defined, and hence the model is
integrable, for all values of the parameters $q$ and $m$
($m$ is a boundary parameter). 
Although the model describes a system of interacting spins in the same
way as the XXZ model \cite{\FT}, it provides a framework for treating the
effect of boundaries which is significantly different from previous
approaches. 
In particular some of the properties of XXZ often invoked in 
implementing the Bethe ansatz 
(such as symmetry properties) 
do not hold in the usual way. 
Then again, 
this model has simple reference states, 
while the usual open chain with nondiagonal boundary does not
(although important progress has been
made towards determining the spectrum of the ordinary open chain recently
\cite{Nepomechie03} despite the lack of obvious reference states). 
We prove here that the ordinary open chain and our new model {\em with} reference
states are equivalent. 
This will allow us to take a line of lower resistance through Bethe
ansatz calculations than either model offers alone. 

The new model turns out to offer a more general treatment of boundaries than
previously possible, so this motivates us to seek
equivalents in this setting for the symmetry  properties of XXZ. 
The first approach to this problem was through algebraic Lie theory
\cite{MartinRyom03}, but there only abstract (although intriguing) 
results about the symmetry of the original extension were found. 
 Here we look directly at the Bethe ansatz, and the most general
possible extension.


Our objectives in {\em  this} paper are firstly to 
put $\Theta$ in a more general setting by considering 
`crossing tensor space representations'
(which ensure a crossing condition necessary for Bethe ansatz); 
then to describe    
extensions of $\Theta$ from $T_n$ to $b_n$ systematically;
then for each of the types of representation of $b_n$ which we find:
(i) to examine the symmetry algebra of the Hamiltonian of 
the resultant spin-chain;
(ii) to investigate the Bethe Ansatz of these Hamiltonians.
The main results in this regard are the symmetries summarized in 
Propositions~\ref{symm q}, \ref{symm q pi} and 
equations 
(\ref{12a}),
(\ref{comh}), (\ref{comhb}), and
the form of the Bethe Ansatz solution in (\ref{BAE}).

In section~\ref{crossX} we discuss the role and implementation of
 crossing symmetry in the algebraic construction. 
In section~\ref{cablingX} we recall the definition of $\Theta$,  
derive the new representations, 
and establish some important notation (swept under the carpet in
\cite{DoikouMartin03}). 
We then derive the universal algebraic Hamiltonian. 
In  section~\ref{qgX} we recall the role of quantum
groups in the symmetry of ordinary spin chains, and discuss how
this might generalize to our case. We find a number of actions of
quantum groups on our chain, but not the complete symmetry algebra. 
We discuss how one might proceed to find the complete symmetry, 
and illustrate some subtleties compared to the XXZ spin chain case \cite{\FT}.
The remainder of the paper is concerned with the solution of 
the Bethe Ansatz for the most physically interesting cases. 


A comment is in order on the physical motivation for this approach.
The aim is to understand and compute with directly physically relevant
models \cite{LiebMattis66}. 
However, 
not every directly physically relevant model is integrable, and those
integrable models, such as XXZ, which do have arguable physical
relevance do not in general remain integrable for arbitrary boundary
conditions (or, if they do, present significant technical problems). 
The idea here is to consider models which {\em are} 
integrable with suitably
general boundary conditions, sacrificing the direct superficial
similarity with XXZ. However we then {\em prove} these models  
to have the same spectrum (up
to multiplicity) as more manifestly physically relevant models. 

\subsection{The blob algebra and the boundary YBE} \label{ss1.1}
\begin{de} 
Let $q$, $\delta_{e}$, $\kappa$ be given scalars, and $\delta=-q-q^{-1}$.
The blob algebra
 $b_n=b_n(q,\delta_{e}, \kappa)$ is defined by generators
$\abU_{1},\abU_{2},...,\abU_{n-1}$ 
and $\abe$, and relations \cite{MartinSaleur94a}:
\begin{eqnarray} 
\abU_{i}\ \abU_{i} &=& \delta\ \abU_{i} \label{TL:UU=U}\\
\abU_{i}\ \abU_{i+1}\ \abU_{i} &=& \abU_{i} \non\\ 
\Big [ \abU_{i},\ \abU_{j} \Big ]  &=& 0, \qquad ~~|i-j| \neq 1 \label{TL} 
\end{eqnarray} (so far we have
the ordinary Temperley--Lieb algebra $T_{n}(q)$ 
\cite{TemperleyLieb71})
\begin{eqnarray}
\abe\ \abe &=& \delta_{e}\ \abe   
\label{delta_e}
\\
\abU_{i}\ \abe\ \abU_{i} &=& \kappa\ \abU_{i}   
\label{kappa}
\\ 
\Big [ \abU_{i},\ \abe \Big ]  &=& 0, \qquad ~~i \neq 1. 
\label{blob}  
\end{eqnarray} 
\end{de}
Note that we are
free to renormalize $\abe$, changing only $\delta_{e}$ and $\kappa$
(by the same factor), thus from $\delta,\delta_{e},\kappa$ there
are really only two relevant parameters 
(originally \cite{DoikouMartin03} this freedom was used to fix $\delta_e =1$, but
this is not generally the best choice). 
It will be natural later
on to reparameterize so that the three are related (they only depend on
$q$ and a single additional parameter $m$), 
but it is convenient to treat them separately
for the moment, and leave $m$ hidden.


Let 
$\rho: T_n(q) \to \End(V^{\otimes n})$ 
be a tensor space representation of $T_n(q)$ (see
section~\ref{crossX}),
${\cal P}$ be the permutation operator on $V \otimes V$,
and $q=e^{i\mu}$. 
Then a solution to  (\ref{YBE}) is given by \cite{Baxter82}
\be 
\label{master}
R_{i\ i+1}(\lambda) =
{\cal P}_{i\ i+1} (\sinh \mu (\lambda+i) \ \rho(1)  
\ + \  \sinh \mu\lambda \ \rho (\abU_{i})).
\eeq

Suppose $\rho$ extends to a representation of $b_n$. 
Then the analogous construction for 
a $K$-matrix solving the reflection equation (\ref{RE}) is
\cite{LevyMartin94,DoikouMartin03}
\be
K(\lambda) = 
x(\lambda)\ \rho(1)  
\ + \  y(\lambda)\ \rho(\abe) 
\label{ansatz1}
\eeq 
where 
\be 
x(\lambda)= -\delta_{e}\cosh \mu (2\lambda +i) -
\kappa \cosh 2\mu \lambda -\cosh 2 i\mu \zeta
 \hspace{1cm}
y(\lambda)=2 \sinh 2\mu\lambda\  \sinh i\mu 
\label{ansatz2} 
\eeq
(here $\zeta$ is an arbitrary constant).

\section{The Temperley-Lieb algebra and tensor space}\label{crossX}

Although we will be concerned with using the blob algebra to 
treat boundary solutions, it is helpful to begin by unpacking a
little what is meant by a {\em tensor space} representation, 
and the crossing symmetry condition. 

\subsection{Preliminaries}
\newcommand{\vb}{{\mbox{b}}}
The next few results are elementary. 
$\;$ 
Let $k$ be a field, $N$ a natural number and $V $ an $N$-dimensional
$k$-vector space with basis $B = \{ \vb_1, \vb_2, \ldots \vb_N \}$. 
The permutation matrix $\Perm{ij}$ acts on $V^{\otimes n}$ by 
\[
\Perm{ij} \; 
v_1 \otimes \ldots \otimes v_i \otimes \ldots  \otimes v_j \otimes \ldots
\; = \; 
v_1 \otimes \ldots \otimes v_j \otimes \ldots  \otimes v_i \otimes \ldots
\qquad (v_{k} \in V)
\]
A standard picture of this is that given a {\em chip} with rows of $n$
legs, the action of $\Perm{ij}$ is to cross the corresponding legs:
\[
\includegraphics{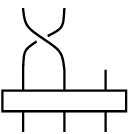}
\]
(here the over/under information is unimportant). 

For $M \in \End(V^{\otimes n})$ the $i^{th}$ factor transpose is
defined by 
\[
\la \ldots v_i' \ldots | \; M^{t_i} \; |  \ldots v_i \ldots \ran
\; = \; 
\la \ldots v_i \ldots | \; M \; |  \ldots v_i' \ldots \ran
\qquad (v_{k} \in B)
\]
(so the usual total transpose is $t=t_1 t_2 \ldots t_n$). 
In the chip realisation, $t_i$ is a kind of {\em $s$-channel crossing}:
\[
\includegraphics{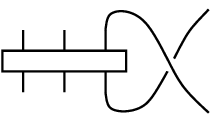}
\]

\prl(Pt)
For any $i,j$
\[
\Perm{ij}^{t_i} = \Perm{ij}^{t_j}
\]
\[
\la v_1' v_2' | \;  \Perm{ij}^{t_j} \; | v_1 v_2 \ran 
= \delta_{v_1 v_2} \delta_{v_1' v_2'}
\]
\end{pr}
The proof is elementary (if slightly tedious). An example is more
enlightening:
{\small \[
\xymatrix{
{\mat{cc|cc} 1&&& \\ &0&1& \\ \hline &1&0& \\ &&&1 \tam}   
\ar@{->}[r]^{t_2} \ar@{->}[d]^{t_1} & 
{\mat{cccc} 1&&&1 \\ &0&0 \\ &0&0 \\ 1&&&1 \tam} 
\ar@{=}[dl]^{}
\\
{\mat{cccc} 1&&&1 \\ &0&0 \\ &0&0 \\ 1&&&1 \tam }
}
\]}
The composite index contraction $\Perm{ij}^{t_j}$ may be represented pictorially by
\[
\includegraphics{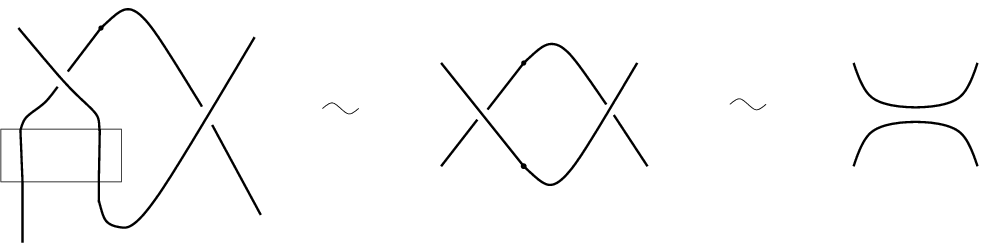}
\]
(NB, this index contraction 
should not be confused with a Temperley-Lieb diagram --- see later). 
A trivial corollary is 
\eql(PPt)
\Perm{ij} \Perm{ij}^{t_j} =  \Perm{ij}^{t_j}
\eq

\newcommand{\cV}{{\mathcal V}}
\prl(vp)
For any $\cV \in \End(V)$
\[
(\cV \otimes \Id) \Perm{12}^{t_1} = (\Id \otimes \cV)^t \Perm{12}^{t_1}
\]
\end{pr}
The proof is a direct calculation. 
The picture for this is 
\[
\includegraphics{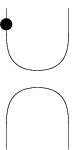}
\qquad \raisebox{21pt}{ = } \qquad
\includegraphics{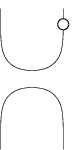}
\]
(i.e. $\cV$ is a black dot and $\cV^t$ a white dot). 

\subsection{Realisations of $T_n(q)$}
\begin{theo}
Provided that
\eql(condition01)
\cV^t \cV^t \cV \cV =  \Id
\eq 
\eql(condition:trace)
\tr(\cV \cV^t) \; =  \; -( q+q^{-1} ) 
\eq
then
\eql(V construct)
\rho_{\cV}(\abU_i) = \cV_{\ij} \Perm{i\;\ij}^{t_{\ij}} \cV_{i}
\eq
is a representation of $T_n(q)$ on $V^{\otimes n}$ for any $n$.   
\end{theo}
{\em Proof:}
Consider the following identities:
\eql(uu=u pic)
\raisebox{-30pt}{ \includegraphics{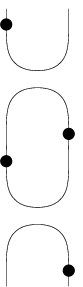}} \qquad
= ( \cV_2 \Perm{12}^{t_2} \cV_1 )^2
=
\;  \tr( \cV \cV^t )  \; \cV_2 \Perm{12}^{t_2} \cV_1
\eq
\eql(uuu=u pic)
\includegraphics{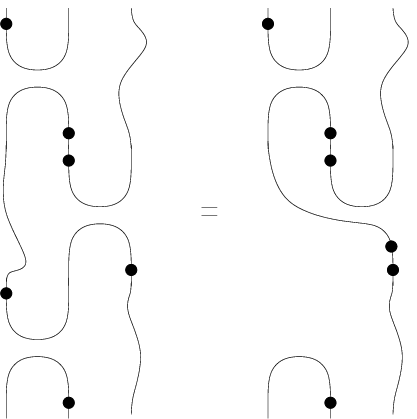}
\eq
\Qed

\noindent
Remark: These pictures are a departure from the standard diagram calculus for the
Temperley-Lieb algebra. In the calculus the lines represent more than a simple
index contraction, and 
decorations are superfluous. 
Here the lines {\em are} merely index contractions 
(so, without the decorations we have only a realisation of 
$T_n(q)$ with $q+q^{-1} = N$, i.e. 
{\em classical} Temperley-Lieb in case $N=2$). 
\footnote{%
These remarks also serve to differentiate between the
  construction here and similar looking diagrams for 
'wreath' Temperley-Lieb algebras such as in \cite{RuiXi03}. 
None-the-less, this decorated form is familiar in some long standing 
diagrammatic treatments of the
Temperley-Lieb algebra, and it has been tacitly assumed that choices
for $\cV$ correspond to variations on the XXZ representation 
(which commutes with the action of $U_q(sl(2))$ on tensor space when
$N=2$)
by a combination of cabling (i.e.  $U_q(sl(2))$ related) and similarity
transformations. 
It turns out that the construction is more general!
(It has not widely been considered likely that there {\em are} tensor space
representations outside the $U_q(sl(2))$  contruction. 
So we should confess that we would have overlooked the fact again, 
were it not for some serendipity in our investigation of the crossing
properties of the representation introduced in \cite{MartinWoodcock03}.)
}


\begin{de}
A representation $\rho$ of $T_n(q)$ is a 
{\em local} tensor space representation if it
acts on the tensor product $V^{\otimes n}$ for some $V$ and 
\[
\rho(\abU_{i+1}) = 
\Perm{i\;\ij}
\Perm{\ij \; i\! +\! 2} \; \rho(\abU_{i}) \; \Perm{\ij \; i\! +\! 2}
\Perm{i\;\ij}
\]
and $\rho(\abU_i)$ acts trivially on all but the $i^{th}$ and $i+1$st
tensor factor. 
\end{de}
A tensor space $R$-matrix 
$R_{ij} = \Perm{\ij \; j} \; R_{i\;\ij} \; \Perm{\ij \; j}$
is said to satisfy the {\em unitarity } condition if 
\eql(unitarity)
R_{ij} (\lambda) R_{ji}(-\lambda) \propto \Id
\eq
and said to satisfy the {\em crossing } condition 
\cite{Sklyanin88,MezincescuNepomechie92b} if 
\eql(crossing1)
\cV_i R_{ij}^{t_j} (\lambda) \cV_i
= -R_{ij}(-\lambda-i)
\eq
for some matrix $\cV \in \End(V)$. 


\begin{theo}
Let  $R_{ij} = \Perm{\ij \; j} \; R_{i\;\ij} \; \Perm{\ij \; j}$
be constructed as in (\ref{master}), 
using  the $\cV$-construction  (\ref{V construct}):
\eql(all in)
R_{ij} (\lambda) \; = \; \sinh(\mu(\lambda+i)) \Perm{ij} 
              \; + \; \sinh(\mu\lambda) \; \Perm{ij} \cV_j \Perm{ij}^{t_j} \cV_i  
\eq
with  $\cV_i^2 = \Id$. 
Then $R_{ij}$ solves the Yang-Baxter equations; 
obeys the unitarity condition; 
and
has the  crossing condition property. 
\footnote{
   Note that  $\cV_i^2 = \Id$  
   implies (\ref{condition01}). 
   Thus sufficient conditions for integrability with parameter $q$ are  
   $\cV_i^2 = \Id$ and the trace condition. 
}

\noindent
(For this reason a representation of form $\rho_{\cV}$ is called a
   {\em crossing} representation.) 


Conversely, suppose that $\rho$ is a tensor space 
representation of $T_n(q)$ on $V^{\otimes n}$ (any $V$), and that 
the crossing
property holds with crossing matrix $\cV$ at $\lambda=0$:
\eql(starboy)
\cV_i R_{i\;\ij}^{t_{\ij}} (0) \cV_i \; =\;  -R_{i\;\ij}(-i)
\eq 
Then
\[
\rho(\abU_i) = \Perm{i\;\ij} \cV_{i} \Perm{i \; \ij}^{t_{\ij}} \cV_{i} 
= \cV_{i+1} \Perm{i \; i+1}^{t_{i+1}} \cV_{i}
\]
In other words $\rho$ can always be considered to be determined by $\cV$.

\end{theo}
{\em Proof:}
The TL relations already  imply a 
solution to YBE --- this is well known (see \S\ref{ss1.1}). 
The solution to the unitarity condition 
is a direct consequence of the TL relation (\ref{TL:UU=U}) in particular.
Now, however, if we use the $\cV$-construction, since $R_{ij}$ is given by
(\ref{all in}) 
we have 
\[
\cV_i R_{ij}^{t_j} (\lambda) \cV_i
=  \sinh(\mu(\lambda+i)) \cV_i \Perm{ij}^{t_j}  \cV_i
 + \sinh(\mu\lambda) \cV_i ( \Perm{ij} \cV_j \Perm{ij}^{t_j} \cV_i )^{t_j} \cV_i
\]
\[
\stackrel{(\ref{PPt})}{=}  
\sinh(\mu(\lambda+i)) \Perm{ij} \cV_j \Perm{ij}^{t_j}  \cV_i
 + \sinh(\mu\lambda) \cV_i^2  \Perm{ij}^{}  \cV_i^2
\]
Now apply  $\cV_i^2 = \Id$ to get (\ref{crossing1}). 
The converse is a routine manipulation after substitution of 
(\ref{all in}) into (\ref{starboy}). 
\Qed

\subsection{Examples}
\newcommand{\ux}{u}%

A classification of solutions to the conditions 
$\cV^2=\Id$ and (\ref{condition:trace}) 
for general $N$ is not
our objective here (although it is an interesting problem). 
We will restrict ourselves to the examples we need.
We are interested in solutions which are valid for a fixed but
arbitrary value of $q$ (or equivalently with $q$ regarded as an
indeterminate). 

Let $\ux$ be such that $\ux^2 = -q$. 
The $\rho_{\cV}$ construction is clearly trivial unless $N \geq 2$.  
In case $N=2$, 
a solution to the conditions is given by 
\eql(zz=0)
   \cV = \mat{cc} 0& \ux \\ \ux^{-1} & 0  \tam
 \qquad  \longrightarrow \qquad
   \cV_2 \Perm{12}^{t_2} \cV_1 = 
\mat{cccc} 0  \\ &-q&1 \\ &1&-q^{-1} \\
   &&&0 \tam
\qquad =: \; U(q)
\eq
That is, with    $V ={\mathbb C}^{2}$, 
\be 
   \rrep{q} ( \abU_{l} ) \; \; 
:= \; \rho_{\cV}  ( \abU_{l} ) \; 
= 1 \otimes 1 \otimes \ldots  \otimes 
U(q)
\otimes \ldots \otimes 1 \otimes 1
\;\;\; \in \;\; End(V^{\nm})
\label{tlg} 
\eeq 
(acting non-trivially  on $V_{l} \otimes V_{l+1}$) 
defines a representation of $T_{\nm}(q)$  (any $q$, $\nm$). 
This is the representation arising in the $\nm$-site XXZ spin chain 
\cite{Baxter82}. 

   More generally, with 
   \[
   \cV = \mat{cc} a&b \\ c&d \tam
   \]
   the square condition gives $a^2=d^2$, $b(a+d)=c(a+d)=0$.
   In case $a=d>0$ then $b=c=0$ so there is no solution 
   to the trace condition for $q$ indeterminate.
\newcommand{\zz}{\alpha}%
\newcommand{\uw}{u}%
In case $a=-d$ then there is a solution 
\eql(cross mix)
   \cV = \mat{cc} \sqrt{-\zz(\zz+\uw+\uw^{-1})} & \zz+\uw \\ 
   \zz+\uw^{-1} & -\sqrt{-\zz(\zz+\uw+\uw^{-1})} \tam
\eq
   for each $\zz$. Note that the solution in (\ref{zz=0}) is the $\zz=0$
   case. 
   Since this space of solutions is continous with (\ref{zz=0}) it
   follows that the representations of $T_n(q)$ constructed are
   generically equivalent. (To see this note that characters will be
   continuous functions of $\zz$, but irreducible multiplicities are
   integers. The only continuous integer valued functions are constants,
   thus the irreducible content of the representation does not depend on
   $\zz$.) 

   We have shown that there is only one $q$-indeterminate class of solutions 
   for $N=2$, and that the representations of $T_n(q)$ arising are all
   generically equivalent. 

   It is worth remarking that (\ref{cross mix}) is the first example of a
   mixed state crossing matrix to be observed. By virtue of our previous
   remark this generalisation does not seem to be of particular intrinsic
   interest. The possibilites for higher $N$, however, are intriguing 
   (but not considered further here). 


   The example which concerns us (because of its role in building
   representations  of $b_n$) has $N=4$. 
   Noting the example above we restrict attention to antidiagonal
   $\cV$. Then we have that 
   $\cV = \mbox{antidiagonal}(a,b,b^{-1},a^{-1})$ and that 
   $a^2 +a^{-2}+b^2+b^{-2} = -(q+q^{-1})$. 
   There are a number of distinct classes of solutions. 
In order to understand the significance of the so called {\em \gemini} solution 
(given explicitly shorly) it
is appropriate to recall its {\em original} construction 
\cite{MartinWoodcock03},
in which the crossing property appears to be just a happy accident. 

\section{The \gemini\
representations and spin chains}\label{cablingX}

Define 
\be 
[x]_{q} = {q^{x} -q^{-x} \over q-q^{-1}} 
= \frac{e^{i\mu x}-e^{-i\mu x}}{e^{i\mu }-e^{-i\mu }} 
= \frac{\sinh(i \mu x)}{\sinh(i \mu)} 
\eeq
Set 
\be 
   r=i\sqrt{iq}, ~~ \qquad  \rs =\sqrt{iq}  
\eeq
so that $ r \rs = -q$. 
The map 
$\Theta: T_{n}(q) \to End(V^{2n})$ is
constructed by combining parts of the representations $\rrep{r}$
of $T_{2n}(r)$ and $\rrep{\rs}$ of $T_{2n}(\rs)$ as follows:
\be
   \Theta(\abU_l)  
   = \rrep{r} ( \abU_{n-l} ) \; \rrep{\rs} ( \abU_{n+l} )
\label{Theta}
\eeq
It is easy to check that this is a representation. 
Because of the way it melds two chains with {\em different} quantum parameters
this is called the \AC\ or \gemini\ representation.


   Note that the Temperley--Lieb algebra is {\it not} a bialgebra, so
   the realization (\ref{Theta}) is not an obvious result. 
The meld gives a new kind of tensor space representation. 
It has been studied from
the point of view of quasihereditary algebras in \cite{MartinRyom03}, 
but given the importance of the orthodox XXZ representation in
integrable physics,
and the properties of $\Theta$ we are about to elucidate, 
this representation merits more concrete study.
(It is easy to see that this gemini idea is amenable to further massive
generalization, but we will not pursue the point here.)

   A consequence of the construction is that $\Theta$ is a crossing
   representation (the crossing matrix is given in (\ref{ACV})). 
Another striking feature of  this representation is that it extends to an 
interesting representation of $b_{n}$ in a number of distinct ways.
It will be evident that $\Theta(\abU_1)$ acts on 
$V^{\otimes 4} = V_{n-1}\otimes V_{n}\otimes V_{n+1}\otimes V_{n+2}$ 
as a rank 1 matrix. Consider for a moment the case $n=2$:
\eql(rank 1 1)
\label{the rank 1 property}
\Theta(\abU_1) = \frac{1}{-q}
((0,-r,1,0) \otimes (0,-\rs,1,0))^t . ((0,-r,1,0) \otimes (0,-\rs,1,0)) 
\label{tht}
\eq
In consequence {\em any} matrix $M$ acting non-trivially only on this
$V^{\otimes 4}$ obeys $\Theta(\abU_1) M \Theta(\abU_1) = k^M \Theta(\abU_1)$ 
for some $k^M$ --- cf. relation~(\ref{kappa}). 
On the other hand relation~(\ref{blob}) is satisfied by 
any matrix $\Theta(\abe)$ acting non-trivially only on the middle
two factors ($V_{n}\otimes V_{n+1}$) of this $V^{\otimes 4}$. 
We will call this the {\em local condition}.  
(NB, this is  not a necessary condition, but our choice here --- 
a consequence of this choice is that the solutions we consider will be
characterised by a 
$4\times 4$ matrix ${\cal M}$ giving the action on $V_n \otimes V_{n+1}$). 
Finally (\ref{delta_e}) requires that the spectrum
of any candidate for $\Theta(\abe)$ be taken from $\{ \delta_e, 0 \}$. 
We do not pursue the most general form for the extension here, but
concentrate on a representative set of interesting choices
(i.e. with characterising matrix  ${\cal M}$ of rank 1 or 2). 

   We have the following justification for this level of generality:
\prl(big one)
The spectrum of the spin chain model 
built from a representation of $b_n$ of the
form of  $\Theta$ depends on  ${\cal M}$  only through the rank of  ${\cal M}$.
\end{pr}
   We will prove this result in section~\ref{rep thy}.

Fixing $q$, define matrices in $\End(V^{\otimes 2n})$ as follows:
\be
M^i(Q) \; = \;\;  
   \frac{-\delta_e}{Q+Q^{-1}}   
   \;\;  1 \otimes 1 \otimes \ldots
   \otimes\left( \begin{array}{cccc}
    0   &0        &0       &0   \\
    0    &-Q      & 1     &0   \\
    0    &1        &-Q^{-1} &0   \\
    0   &0        &0       &0
   \end{array} \right) \otimes \ldots \otimes 1 \otimes 1 \label{cabl} 
\eeq 
   where the $4\times 4$ matrix acts on $V_{n} \otimes V_{n+1}$ 
   and $Q$ is some scalar; 
   \be  
   M^{ii}(Q) =  
   \frac{-\delta_e}{Q+Q^{-1}}   
   1 \otimes 1 \otimes \ldots \otimes\left( \begin{array}{cccc}
    -Q    &0        &0       &1   \\
    0    &0      & 0     &0   \\
    0    &0        &0 &0   \\
    1    &0        &0       &- Q^{-1}
   \end{array} \right) \otimes \ldots \otimes 1 \otimes 1
   \label{cabl2} 
   \eeq 
\be 
   M^{+}(Q,Q') = M^{i}(Q) + M^{ii}(Q') \label{++}
\eeq
\be  
M^{iii}(Q_1,Q_2) =  
\frac{\delta_e}{(Q_{1}+Q^{-1}_{1})(Q_{2}+Q^{-1}_{2})}  
   \;  1 \otimes 1 \otimes \ldots \otimes \left(
   \begin{array}{cc}
    -Q_1    &1          \\
    1    &-Q_1^{-1}
   \end{array} \right) \otimes \left(
   \begin{array}{cc}
    -Q_2^{-1}      &1          \\
    1    &-Q_2
   \end{array} \right) \otimes \ldots \otimes 1 \otimes 1
\label{cabl3} 
\eeq 
where the $2 \times 2$  
matrices act separately on $V_{n}$, $V_{n+1}$ respectively; 
and $M^{iii}(Q)=M^{iii}(  i\sqrt{iQ}  ,   \sqrt{iQ} )$.

For $I \in \{i,ii,+,iii \}$  direct calculation shows
\[
\Theta(\abU_1) \; M^{I} \;  \Theta(\abU_1) \; = \; \kappa^I \; \Theta(\abU_1)
\]
where
\be 
\label{kap}
\kappa^{i}(Q)  = {-\delta_e ( q^{-1}Q+qQ^{-1} ) \over Q+Q^{-1} } , 
\qquad
\kappa^{ii}(Q) = {-\delta_e ( i Q -i Q^{-1} ) \over  Q+Q^{-1} } , 
\eeq 
\be 
\label{kapp}
\kappa^{+}(Q,Q') = \kappa^{i}(Q) + \kappa^{ii}(Q') , 
\qquad
\kappa^{iii}(Q) = {-\delta_e ( q^{-1}Q+qQ^{-1}+2 ) \over Q+Q^{-1} } 
\eeq 
so 
\prl(theta reps)
For each $I \in \{i,ii,+,iii \}$ there is a representation 
   $\Theta^I : b_n(q,\delta_e,\kappa) \rightarrow End(V^{\otimes 2n})$ 
   given by 
   $\Theta^I(\abU_i)=\Theta(\abU_i)$, 
$\Theta^I(\abe)=M^I$, 
provided 
$\kappa^I(Q) = \kappa$. 
(NB, for given $I$ this is a condition on $Q$.%
\footnote{For example, in case $i$ the condition is 
$Q^2 = -\frac{\kappa+\delta_e q}{\kappa+\delta_e q^{-1}}$
and in case $ii$ it is 
$Q^2 = -\frac{\kappa -i \delta_e }{\kappa+ i \delta_e }$.})
\end{pr}
The representation $\Theta^{i}$ is the representation $\rho_{0}$ in
\cite{MartinWoodcock03}; 
while $\Theta^+$ with $Q' \rightarrow 0$ is  the representation $\rho_{x}$
there. 
The other representations are new.

\smallskip


We will need to have to hand one more kind of representation of $b_n$.
It will be evident that if $\kappa = 0$ then the quotient relation
$\abe=0$ is consistent with the defining relations of $b_n$.
It follows that for this particular value of $\kappa$ 
any representation of $T_n$ may be extended to a representation of
$b_n$ by representing $\abe$ by zero. 
In case we start with $\Theta$, we will write $\Theta^0$ for this
extension to $b_n$. 
It will be evident from (\ref{ansatz1}) that this gives a trivial $K$-matrix.

\subsection{Anatomy of a spin chain}
   \newcommand{\lef}[1]{{#1}^-}
   \newcommand{\ri}[1]{{#1}^+}

It will be convenient to be able to refer to individual factors in
the tensor product $V^{\otimes N}$. Normally we write
$
   V^{\otimes N} = V_1 \otimes V_2 \otimes \ldots \otimes V_N
$
so $V_{i}$ is the space of the $i^{th}$ spin in the spin chain. In
our situation, however, with $N=2n$, it will be convenient to relabel
\[
V^{\otimes 2n} = V_{\lef{n}} \otimes V_{\lef{(n-1)}} \otimes \ldots V_{\lef1} 
   \otimes V_{\ri1}
   \otimes \ldots \otimes V_{\ri{(n-1)}} \otimes V_{\ri n} .
\]
\begin{figure}
\[
\includegraphics{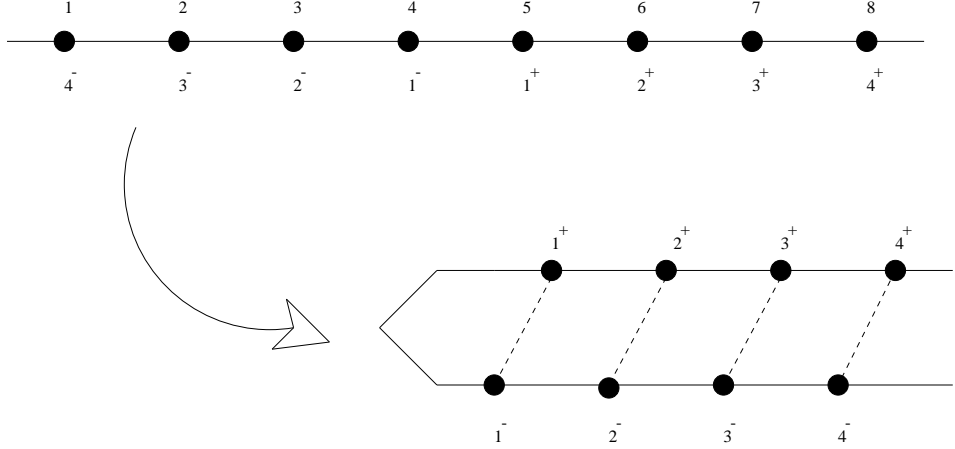}
\]
\caption{\label{spinchain} Conventional spin chain site numbering,
and mirror image pair site numbering. (The new composite sites are
shown clustered by dotted lines.)
NB, Here the chain links merely induce an order on the vertices,
convenient for labelling --- they do not correspond to tensor index
contractions.}
\end{figure}
I.e.  
\be 
n+l \to \ri l, ~~~ \qquad  n-l+1 \to \lef l, ~~ \qquad l=1,\ldots,n. 
\label{iden} 
\eeq 
One may think of this as
a product of a mirror image pair of factors $V^{\otimes n}$, as the new
labelling suggests. One should then think of `folding up' the
linear chain at the point between $V_{\lef 1}$ and $V_{\ri 1}$, so that it
becomes a double thickness chain with composite sites of form
$V_{\lef i} \otimes V_{\ri i}$. Thus in particular the composite site
$V_{\lef1} \otimes V_{\ri1}$ lies at one {\em end} of the chain, 
and $\abe$ acts non-trivially only on this boundary site. 
All of these points are illustrated by figure~\ref{spinchain}.

   \newcommand{\Vs}{{\bf V}}
   \newcommand{\Iss}{{\mathcal I}}

This labelling issue is mundane, but it is practically significant here.
Let $\Vs = (V^{\otimes N}, (l_1,l_2,\ldots,l_N))$ be a tensor power of a
vector space $V$ together with  a labelling scheme for the tensor
factors.
Then for each $M$ acting on $V^{\otimes m}$ ($m\leq N$) and list 
$\{ i_1, i_2, \ldots, i_m \} \subseteq \{ l_1, l_2, \ldots, l_N \}$
we define 
$\Iss^{\Vs}_{i_1, i_2, \ldots, i_m}(M)$ to be  the 
matrix acting on $V^{\otimes n}$ 
which acts on the subfactors $V_{i_1} \otimes V_{i_2} ...$ as $M$, and
acts trivially on any other factors.
It will be convenient to use the $R$-index  (or tensor space index) 
notation for actions on tensor space
\cite{DoikouMartin03}, 
in which, given 
$\Vs$, we write simply $M_{i_1 i_2 \ldots i_m}$ for 
$\Iss^{\Vs}_{i_1, i_2, \ldots, i_m}(M)$. 
Thus 
$
\rrep{q}(\abU_i) = \left( U(q)  \right)_{i \; i\!+\!1}
$
for general $n$, for the ordinary labelling scheme of $V^{\otimes n}$. 
However, after the `folding' (\ref{iden}) of the open spin chain, 
the tensor factors have been relabelled, so the $R$-index notation
there gives: 
\be 
{\cal R}_{\rs}(\abU_{n+l}) = (U(\rs))_{\ri l \; \ri{(l\!+\!1)}}, ~~ \qquad
   {\cal R}_{ r}(\abU_{n-l}) 
      = (U(r^{}))_{\lef{(l\!+\!1)} \; \lef l}   
      = (U(r^{-1}))_{\lef l \; \lef{(l\!+\!1)}} , 
\label{iden2} 
\eeq 
so
\be
\Theta(\abU_l) = 
(U(r^{-1}))_{\lef l \; \lef{(l+1)}} \;\; (U(\rs))_{\ri l \;  \ri{(l+1)}} 
\label{the}
\eeq  


Notice from (\ref{Theta}) that the single index $l$ on $\abU_l$ is
associated to a mirror image pair $(\lef l,\ri l)$ in the underlying 
$V^{\otimes 2n}$
(a coupled pair in the folded scheme).
Accordingly we introduce the space/mirror--space notation 
\newcommand{\LLL}{^-}
\newcommand{\RRR}{^+}
$\tilde l = (\lef l, \ri l)$. 
We correspondingly extend the tensor space index notation so that 
if an operator $M$ acts on ${V}^{\otimes m}$ then $M_{\tilde i_i \tilde i_2 ...}$
acts on $V^{\otimes 2n}$ as 
$M_{\lef{i_1} \lef{i_2} ... }  M_{\ri{i_1} \ri{i_2} ... }$.
Thus ${\cal P}_{\tilde k \tilde l}
        ={\cal P}_{\lef k \lef l}\ {\cal P}_{\ri k \ri l}$.
While if $M$ acts on $({V\otimes V})^{\otimes m}$ then 
$M_{\tilde i_i \tilde i_2 ...}$
acts on $V^{\otimes 2n}$ by acting 
non-trivially on the appropriately positioned mirror image pairs: 
$(V_{\lef{i_1}} \otimes V_{\ri{i_1}})
\otimes (V_{\lef{i_2}} \otimes V_{\ri{i_2}})\otimes ...$. 

\subsection{$R$-matrices and $K$-matrices for the gemini spin chain}

The gemini spin chain is 
the solution to the Yang-Baxter equation (\ref{YBE}) 
corresponding to using $\Theta$ in (\ref{master}). 
The $R$-matrix can be written using the index notation as follows. 
Define
$
   \check U_{k l}(r)= {\cal P}_{kl}\ (U(r))_{k l}  
$ 
and for $k \neq l \in \{1,2,...,n\}$
\be 
   R_{\tilde k \tilde l}(\lambda)
   = {\cal P}_{\lef k \lef l}\ {\cal P}_{\ri k \ri l} 
   \left(  \sinh \mu(\lambda+i)\ + \sinh\mu \lambda\  
    U_{\lef k  \lef l}(r^{-1})\  U_{\ri k  \ri l}(\rs) \right)
   \\
   = \sinh \mu(\lambda+i)\ {\cal P}_{\lef k \lef l}\ {\cal P}_{\ri k  \ri l}
   + \sinh\mu \lambda\  
   \check U_{\lef k   \lef l}(r^{-1})\ \check U_{\ri k   \ri l}(\rs) 
\label{R2} 
\eeq 
For $n=2$ this is a $16 \times 16 $ matrix, given explicitly in Appendix A.
Define  $\check R_{\tilde k \tilde l}(\lambda) =
{\cal P}_{\tilde k \tilde l}\ R_{\tilde k \tilde l}(\lambda)$. 

Note that the bulk
space is significantly changed from that of the basic YBE solution
for XXZ. {\em Here} the entire bulk space acquires a mirror image
(a mirror copy $V_{\ri l}$ of each $V_{\lef l}$). 
Neither side, viewed in isolation, retains the
defining $q$--parameter in the usual way, 
nor do they have the same $q$--parameter
as each other,%
\footnote{This system is also radically different from
spin ladder systems such as in 
\cite{Wang99,WangSchlottmann00,TonelFoerster01} 
--- see  \cite{DoikouMartin03}
for an explicit comparison; and from systems obtained by fusion.}
and $K$ acts on $V_{\lef 1} \otimes V_{\ri{1}}$. 


\prl(theta crossing)
Representation $\Theta$ is a crossing representation, with $V= V_- \otimes V_+$ 
($V_{\pm} \cong \C^2$). 
\end{pr}
{\em Proof:}
This $R$--matrix satisfies the unitarity and crossing properties
in the form 
\be 
R_{\tilde k \tilde l}(\lambda)\ R_{\tilde l \tilde k}(-\lambda) \propto \Id,
~~ \qquad
R_{\tilde k \tilde l}(\lambda)
= - {\cal V}_{\tilde k}\ 
R_{\tilde k \tilde l}^{t_{\tilde l}}(-\lambda-i)\ {\cal V}_{\tilde k},
\eeq 
where
\be 
\label{ACV}
{\cal V}_{\tilde k}
   = {\cal V}^X_{\lef k}(r^{-1})\ {\cal V}^X_{\ri k}(\rs),
~~\mbox{and}  ~~{\cal V}^X(p)=\left(
\begin{array}{cc}
                               0                  &  -ip^{-{1\over 2}}  \\
                              ip^{{1\over 2}}  & 0   \\
\end{array} \right)\,. 
\eeq
\Qed

The $K$--matrix (\ref{ansatz1})  
in type $I$ may be given in the $4\times 4$ matrix form 
acting on $V_{\lef 1}\otimes V_{\ri 1}$ as 
$K(\lambda) = \KI{I}(\lambda) = x(\lambda) \Id + y(\lambda) M^I_{n=1}$: 
\be
\KI{i} (\lambda) = \mat{cccc}
x(\lambda) &                &                                     & \\
           & w^{+}(\lambda) &f(\lambda)    \\
           & f(\lambda)  & w^{-}(\lambda)   \\
           &                &                       & x(\lambda)
\tam \qquad
\KI{ii} (\lambda) = \mat{cccc}
w^{+}(\lambda) &     &          & f(\lambda)   \\
           & x(\lambda)         &                                     & \\
           &                &      x(\lambda)         &            \\
f(\lambda)  &   &             &    w^{-}(\lambda) 
\tam
\label{K3} 
\label{K3b} 
\eeq
where 
\be 
   && w^{\pm} (\lambda)
   = x(\lambda) +  {Q^{\pm 1}\delta_{e} \over Q+Q^{-1}}\ y(\lambda) ,
~~~ 
f(\lambda)= -{\delta_{e} \over Q+Q^{-1} }\ y(\lambda)  
\label{ax}  
\eeq 
(We omit the analogous $K$-matrix for types (+),($iii$) for brevity.)
All the $K$-matrices  
satisfy unitarity, i.e.,
\be
K(\lambda)\ K(-\lambda) \propto \Id . 
\eeq

\begin{figure}
\[
\includegraphics[width=7cm]{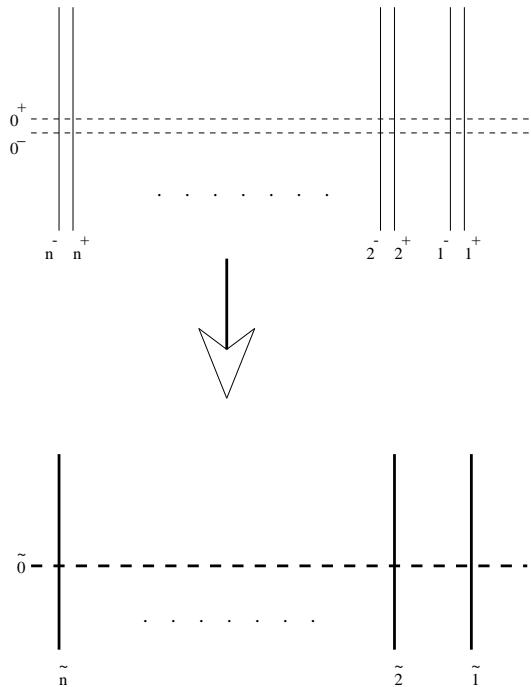}
\]
\caption{\label{mono} The monodromy matrix schematic.
}
\end{figure}

Recall that each spin in a spin chain 
may be thought of as having two legs `in' and
two legs out (the vertex model picture), 
and the factors $V$ in tensor space are the
configuration spaces of individual legs. Since our spins are doubled
up (mirror image pairs) their legs are all composites of two simple legs.
In the usual monodromy matrix formulation  \cite{\FT} of a spin chain 
the legs in the lateral direction (within the transfer matrix layer)
are all labelled 0, while the transverse legs are given the label of
the corresponding spin. 
In our case the lateral direction legs are still composite legs, 
with one component leg coming from the `real' and one from the
`mirror' side. 
Accordingly we will (ab)use the $\tilde l$ notation, so that these
legs are labelled $\tilde 0  = (\lef 0, \ri 0)$
(see figure~\ref{mono})
when it is convenient to do so. 
\subsection{The algebraic Hamiltonian}   
\newcommand{\xtilde}{}
\newcommand{\xR}[2]{R_{\xtilde #1 \xtilde #2}}

Irrespective of whether the indices are composite or otherwise,
the $n$-site chain {\em monodromy matrix} \cite{\FT} is, as usual,
\be 
T_{\xtilde 0}(\lambda) 
= \xR{0}{n}(\lambda) 
\xR{0}{n-1}(\lambda)  \dots  
\xR{0}{1}(\lambda)    ,
\  ~~~ \qquad
\hat T_{\xtilde 0}(\lambda) =
\xR{1}{0}(\lambda) 
\xR{2}{0}(\lambda)   \dots  
\xR{n}{0}(\lambda) 
\label{mono0} 
\eeq
Then by Theorem~1 and Proposition~3 of 
\cite{Sklyanin88} 
the open spin chain {\em transfer matrix} 
\be 
t(\lambda)= tr_{\xtilde 0} M_{\xtilde 0}\
K_{\xtilde 0}^{+}(\lambda)\ T_{\xtilde 0}(\lambda)\ 
K_{\xtilde 0}^{-}(\lambda)\ \hat T_{\xtilde 0}(\lambda), 
\label{transfer} 
\eeq
obeys $[t(\lambda),t(\lambda')]=0$ for all $\lambda,\lambda'$. 
Here  $K^{\pm}_{\xtilde 0}$ denotes the $K$-matrix for the 
left/right boundary of the chain
(and $M_{\xtilde 0}$ is a physically unimportant 
correction to the right boundary
term --- see below).  
In what follows $K^+$ will be unit,  $K^{-}$ 
will be unit or given by 
$K$ from (\ref{ansatz1}) with $\rho$ a crossing representation 
and
$
M_{\xtilde 0}= {\cal V}_{\xtilde 0}^{t}\ {\cal V}_{\xtilde 0} 
$
(its explicit expression in case $\rho=\Theta$ is given in
\ref{A}).  

Note that for $K^{-} = \Id \;$ in particular 
the local $R$-matrix crossing condition (\ref{crossing1})
implies \cite{MezincescuNepomechie92a}
that the open transfer matrix has a crossing symmetry:
\be
\label{cross}
t(\lambda) = t(-\lambda -i) 
\eeq
 
\newcommand{\trace}{{tr}}%
\newcommand{\trac}{\trace_{\xtilde 0}}%
\newcommand{\xwidetilde}{}
\newcommand{\mR}[2]{R_{\xwidetilde{#1} \xwidetilde{#2}}} 
\newcommand{\mP}[2]{{\cal P}_{\xwidetilde{#1} \xwidetilde{#2}}} 
\newcommand{\mRO}[2]{\mP{#1}{#2}} 
\newcommand{\mK}{{K^{-}}} 
\newcommand{\mH}[2]{{\cal H}_{\xwidetilde{#1} \xwidetilde{#2}}} 

Given the set of commuting objects $t(\lambda)$, 
the {\em Hamiltonian} is (up to a choice of overall factor) $t'(0)$ 
   (this commutes with $t(\lambda)$ by elementary considerations). 
Our choice of normalization is 
\be 
{\cal H} 
 = -{\sinh^{1-2n}(i\mu) \over {4\mu x(0)}}
 \Big( \trace_{\xtilde 0} M_{\xtilde 0} \Big )^{-1}\ 
\ t'(0)
\label{H0}
\eeq  
\prl(moochoo)
If 
$R$, $K$ are given as in (\ref{master}), (\ref{ansatz1}), and 
$\rho$ is a crossing representation, then 
the Hamiltonian (\ref{H0}) may be written 
\be 
{\cal H} = {\cal H}^{\rho}   
= \frac{-1}{2}  \sum_{i=1}^{n-1} \rho(\abU_{i}) 
- \frac{\sinh(i\mu) y'(0)}{4\mu x(0)} \rho(\abe)
+ \left( c - \frac{\sinh(i\mu) x'(0)}{4\mu x(0)} \right) \rho(1)
\label{ht} 
\eeq 
where 
$c = -\frac{n}{2}\cosh(\mu i) 
     + \frac{1}{4\cosh(\mu i)}$.
\end{pr}
{\em Proof:} 
   We have from (\ref{master}), (\ref{ansatz1})
   \[
   \mR{k}{l}(0) = \; \sinh(\mu i) \; \mP{k}{l}  \qquad \qquad 
   K^-(0) = \KI{I}(0) = x(0) \Id
   \]
   so
   \begin{eqnarray*}
   t'(0) & = & \trac \left( M_{\xtilde 0} \left(\sinh(\mu i)\right)^{2n-1} x(0)
   \left( 
     \sum_{i=1}^n  \mRO{0}{n} \ldots \mR{0}{i}'(0) \ldots \mRO{0}{1} \Id 
   \mRO{1}{0} \ldots \mRO{i}{0} \ldots \mRO{n}{0}
   \right. \right. \\ && \left. \left.
   +  \frac{\sinh(\mu i)}{x(0)} 
   \mRO{0}{n} \ldots \mRO{0}{2} \mRO{0}{1} \mK_{\xtilde 0}'(0) 
   \mRO{1}{0} \mRO{2}{0} \ldots \mRO{n}{0}
   + \sum_{i=1}^n  \mRO{0}{n} \ldots \mRO{0}{i} \ldots \mRO{0}{1} \Id 
   \mRO{1}{0} \ldots \mR{i}{0}'(0) \ldots \mRO{n}{0}
   \right) \right)
   \end{eqnarray*}
   \begin{eqnarray*}
   &  = & \trac \left( M_{\xtilde 0} \left(\sinh(\mu i)\right)^{2n-1} x(0)
     \left( 
     \sum_{i=1}^n  \mRO{0}{n} \ldots \mRO{0}{i+1} 
   \mR{0}{i}'(0) \mRO{i}{0} \mRO{i+1}{0} \ldots \mRO{n}{0}
   \right. \right. \\ && \left. \left.  \hspace{4cm}
   + \frac{\sinh(\mu i)}{x(0)}   \mRO{0}{1} \mK_{\xtilde 0}'(0) \mRO{1}{0} 
   + \sum_{i=1}^n  \mRO{0}{n} \ldots \mRO{0}{i} \mR{i}{0}'(0) \ldots \mRO{n}{0}
   \right) \right)
   \end{eqnarray*}
   \begin{eqnarray*}
   &  = & \trac \left( M_{\xtilde 0} \left(\sinh(\mu i)\right)^{2n-1} x(0)
     \left( 
   \frac{\sinh(\mu i)}{x(0)}    \mK_{\xtilde 1}'(0) 
   + 2 \sum_{i=1}^{n-1}  \mRO{i+1}{i} \mR{i}{i+1}'(0) 
   + 2 \mRO{0}{n} \mR{n}{0}'(0) 
   \right) \right)
   \end{eqnarray*}
   \begin{eqnarray*}
   &  = & \left(\sinh(\mu i)\right)^{2n-1} x(0) \left( 
   \trac \left( M_{\xtilde 0} \right)
   \left( 
   \frac{\sinh(\mu i)}{x(0)}    \mK_{\xtilde 1}'(0) 
   + 2 \sum_{i=1}^{n-1}  \mRO{i+1}{i} \mR{i}{i+1}'(0) 
   \right)
   +  
   \trac \left( M_{\xtilde 0}  \left( 
   2 \mRO{0}{n} \mR{n}{0}'(0) 
   \right) \right) \right)
   \end{eqnarray*}
   Defining 
   \be 
   \mH{k}{l} =            
   -{1\over 2 \mu} \left( {d\over d \lambda}
   \mP{k}{l}              
   \ \mR{k}{l}(\lambda)   
   \right) \Big \vert_{\lambda =0} 
   \label{H1}
   \eeq 
   we have
   \be {\cal H} =
   \sum_{l=1}^{n-1} \mH{l}{l+1} 
    - {\sinh(i\mu) \over 4 \mu x(0)}  \left( 
   {d \over d \lambda} K_{\xtilde 1}^{-}(\lambda) \right) \Big \vert_{\lambda =0}
    + {\trace_{\xtilde 0} M_{\xtilde 0} \mH{n}{0} 
        \over \trace_{\xtilde 0} M_{\xtilde 0}} 
   \label{H}
   \eeq 
It follows immediately from (\ref{master}) and (\ref{H1}) that 
\be 
\mH{l}{l+1} 
  \; = \; -{1\over 2} \left(  \rho(\abU_l)  
          + \cosh(i\mu) \; \rho(1) \right)
\label{H1'}
\eeq 
and by (\ref{ansatz1}), 
$K'(0) = \;  x'(0) \; \rho(1) \; + \; y'(0) \; \rho(\abe)$. 
Finally by the crossing assumption
\[
\trace_{\xtilde 0} ( M_{\xtilde 0} \mH{n}{0} )  
= \frac{-1}{2} \trace_{\xtilde 0} ( M_{\xtilde 0} 
(\cosh(i\mu)+\cV_0 \Perm{n0}^{t_0} \cV_n ))  
=  \frac{-1}{2} ( \cosh(i\mu) \trace_{\xtilde 0} ( M_{\xtilde 0} ) 
\; + \; \trace_{\xtilde 0} ( M_{\xtilde 0} \cV_0 \Perm{n0}^{t_0} ) \cV_n )  
\]
\[
\stackrel{\mbox{prop.\ref{vp}}}{=}   \frac{-1}{2} ( 
-2\cosh^2(i\mu) \Id
\; + \;  \cV_n^t \cV_n^t \cV_n \trace_{\xtilde 0} (\Perm{n0}^{t_0} ) \cV_n )  
\stackrel{\mbox{prop.\ref{Pt}}}{=}   \frac{-1}{2} (
-2\cosh^2(i\mu) \Id
\; + \;  \cV_n^t \cV_n^t \cV_n  \cV_n  ) 
\]
But $\cV_n^t \cV_n^t \cV_n  \cV_n = \Id$ again by assumption, so we
are done. 




\newcommand{\Gq}[1]{{\cal G}_{#1}}
\newcommand{\cF}{{\cal F}}
\newcommand{\cE}{{\cal E}}%
\newcommand{\cK}{{\cal K}}%

\section{Quantum group symmetry}\label{qgX}
At its most general, the notion of `spin chain Hamiltonian' 
means, amongst other things, a {\em sequence} of
matrices indexed by ${\mathbb N}$, such that the $n^{th}$ matrix
acts on $V^{\otimes n}$ for some $V$ 
(of course only very special kinds of matrices are allowed,
but this structure will be sufficient for now). 
Note that the XXZ Hamiltonian is such a sequence:
\be \label{XXZ H}
\left\{
{\cal H}_{XXZ}
\;\propto\; \sum_{i=1}^{n-1}{\cal R}_{q}(\abU_{i})
   \; = \;  \sum_{i=1}^{n-1} (U(q))_{i \; i+1} 
\hspace{.1in} | \; n=2,3,\ldots \right\}.
\eeq
   Let ${\cal G}$
   be a quantum group \cite{Joseph95} with coproduct    
   $\Delta: {\cal G} \to {\cal G} \otimes {\cal G}$.
   Given a representation
   of ${\cal G} \otimes {\cal G}$ one may use the coproduct to give
   an action of ${\cal G}$. Thus in particular there is a notion of
   tensor product of any two representations of ${\cal G}$ (just as
   there is for ordinary groups).
   By a ${\cal G}$-symmetry of the
   Hamiltonian we mean an action $P$ of ${\cal G}$ on $V$, such that
   the $n$-fold product of $P$ on $V^{\otimes n}$ commutes with
   the $n^{th}$ ${\cal H}$ matrix.
      
In this section we consider the  symmetry of the Hamiltonian
   ${\mathcal H}$ from (\ref{ht}) in cases $\rho=\Theta^I$.
   (Note that at very least a ${\cal G}$-symmetry implies some sort of global limit
   structure on the Hamiltonian sequence.
   Our representation theoretic construction will be used to exhibit such
   a structure in section~\ref{rep thy}.)
   First, it is useful to recall some basic definitions and results.

If  ${\cal G}$ is a bialgebra with coproduct $\Delta$ then 
set $\Delta^2=\Delta$ and then, for $n>2$, 
$\Delta^n 
= (\Delta \otimes 1 \otimes 1 \otimes \ldots \otimes 1) \Delta^{n-1}$
\cite{Joseph95}. 
   The following result is very familiar --- we state it in a slightly
   unusual form suitable for use here.
\prl(commute n)
   Fix a space $V$ and a biaglebra ${\cal G}$ with generators
$\{ {\cal E}_{\alpha}, {\cal F}_{\alpha} , 
    {\cal K}_{\alpha},  {\cal K}_{\alpha}^{-1} \}_{{\alpha} \in I}$
   ($I$ some index set)
   and coproduct
\be
   \Delta({\cal E_{\alpha}}) &=&  {\cal K_{\alpha}}^{-1}
   \otimes{\cal  E_{\alpha}} + {\cal E_{\alpha}} \otimes {\cal
     K_{\alpha}}, \non
   \\
   \Delta({\cal F_{\alpha}}) &=&  {\cal K_{\alpha}}^{-1} \otimes {\cal
     F_{\alpha}} + {\cal F_{\alpha}} \otimes {\cal K_{\alpha}}, \non
   \\
   \Delta({\cal K_{\alpha}}^{\pm 1}) &=& {\cal
   K_{\alpha}}^{\pm 1} \otimes {\cal K_{\alpha}}^{\pm 1}.
   \label{cop alpha}
   \label{cop}
\eeq
   Suppose that $\rho:{\cal G} \rightarrow \End(V)$ a representation,
   so that $\rho^{\otimes 2}(\Delta(-))$ a representation on
   $V_1 \otimes V_2$.
Now let $M \in \End(V_1 \otimes V_2)$ and $K \in \End(V_1 )$
be matrices such that
\be \label{1x}
 [\rho^{\otimes 2}(\Delta(x)),M]=0 , 
\qquad 
[\rho(x),K]=0 
\qquad 
\forall x \in S = \{ {\cal E}_{\alpha}, 
    {\cal K}_{\alpha},  {\cal K}_{\alpha}^{-1} \}.
\eeq
Then with 
   $M_{i \; i+1} = \Iss^{\Vs}_{i \; i+1}(M)$ and $K_{i}=\Iss^{\Vs}_{i }(K)$
\be \label{2x}
[\rho^{\otimes n}(\Delta^n(x)),M_{i \; i+1} ]=0, 
\qquad   ~~~
[\rho^{\otimes n}(\Delta^n(x)),\  K_{i}]=0   
\eeq
for all $n \geq 2$, all $x \in S$, and all $i=1,2,...,n-1$.
\end{pr}
{\em Proof:}
   Firstly, noting that
   $\rho^{\otimes n}(\Delta^n({\cal K}_{\alpha}))
   = (\rho({\cal K}_{\alpha}))^{\otimes n}$, 
for any given $i$
   (\ref{2x})$_{x={\cal K}_{\alpha}^{\pm 1}}$ is
   a direct consequence of the {\em form} of $M_{i \; i+1}$
(respectively $K_i$) 
   and  (\ref{1x})$_{x={\cal K}_{\alpha}^{\pm 1}}$.
   Now define
   \[
   {\chi}_j = \rho^{\otimes n}({\cal K}^{-1}_{\alpha}\otimes \ldots
   \otimes {\cal K}^{-1}_{\alpha}
   \otimes \underbrace{ {\cal E}_{\alpha} }_{\mbox{$j$-th position}}
   \otimes {\cal K}_{\alpha} \ldots \otimes {\cal K}_{\alpha})
   \]
   so that $\rho^{\otimes n}(\Delta^n( {\cal E}_{\alpha} )) = \sum_j \chi_j $.
   If $i>j$ or $i+1 <j$, $\;$ $[\chi_j, M_{i \; i+1} ] = 0$ by
   (\ref{1x})$_{x={\cal K}_{\alpha}^{\pm 1}}$,
while $[\chi_{i} + \chi_{i+1}, M_{i \; i+1} ] = 0$ by
(\ref{1x})$_{x={\cal E_{\alpha}^{}}}$, 
   and so on. 
Similarly $[\chi_{j},\ K_{i}]=0$.  
\Qed

   Comparing Proposition~\ref{commute n} and (\ref{XXZ H}) we see, that for
   Hamiltonians of this form, verification of ${\cal G}$-symmetry may be
   possible by strictly local calculations.

\subsection{Example: XXZ and the quantum group
            $\Uqsl2$}

\begin{de}{\rm \cite{Jimbo85a,Drinfeld86}}
For $q \in \C \setminus \{0,1,-1 \}$ 
let
$ \Gq{}=\Gq{q}  \; := \;U_{q}(sl(2)) $, 
the algebra with generators ${\cal E}$, ${\cal F}$ and ${\cal K}$
and relations
\be
\label{qg2}
\Big [{\cal E},\ {\cal F}\Big ] = {{\cal K}^{2} -{\cal K}^{-2}
  \over q-q^{-1}} , 
\qquad 
{\cal K}\ {\cal E} = q\ {\cal E}\ {\cal K},
&& 
{\cal K}\ {\cal F} = q^{- 1}\ {\cal F}\ {\cal K}. 
\eeq
\end{de}
\begin{lem}
There is an algebra automorphism $w$ on ${\cal G}$ given by 
$w(\cE)=\cF$, $w(\cF)=\cE$, $w(\cK)=\cK^{-1}$
(the `Cartan' automorphism).
\end{lem}
Recall that the standard equivalence classes of
generically simple modules of $ {\cal G}$ may be indexed by the non-negative
integers \cite{Kassel95},
with the module indexed by $\nu$ having dimension $\nu+1$.
In particular we have a $\nu=0$ action on $\C$ given by:
\be
\epsilon({\cal E}) =\epsilon({\cal F})=0,
 ~~\epsilon({\cal K}^{\pm 1})=1
\eeq
and a $\nu=1$ action on ${\mathbb C}^2$ (with basis $\{ v_1, v_2 \}$)
written in terms of Pauli matrices as
\be
\rho({\cal K})=q^{{1\over 2} \sigma^{z}}
= \mat{cc} q^{1/2} & 0 \\ 0 & q^{-1/2} \tam,
~~\rho({\cal E})=\sigma^{+} = \mat{cc} 0&1 \\ 0&0 \tam,
~~\rho({\cal F})=\sigma^{-} = \mat{cc} 0&0 \\ 1&0 \tam.
\label{comre}
\eeq

A simple calculation%
\footnote{
It should be noted that since we have given the representation of
$T_n$ in an explicit matrix form, there must be a consistent convention
for the rendering of direct products into matrix form.
Here we use the `Greek' convention:
\[
(a,b)\otimes (c,d) = (ac, ad, bc,bd) .
\]
}
shows that the $T_{n}(q)$
generators in the XXZ representation (\ref{tlg}) commute with  the
$\rho^{\otimes n}$ action of $U_{q}(sl(2))$ on 
$({\mathbb  C}^{2})^{\otimes n}$
for $n=2$ and hence, via Proposition~\ref{commute n}, any $n$:
\be
\Big [ \rho^{\otimes n} (\Delta^n(x)),\ {\cal R}_{q}(\abU_{i}) \Big ] = 
\Big [ \rho^{\otimes n} (\Delta^n(x)),\ (U(q))_{i \; i\!+\! 1} \Big ] 
=0
\qquad \forall x\in {\cal G}
\label{comre1}
\eeq
That is, we have  the following extraordinary (but very well known)
result.
\newline
The action $\rrep{q}$ of $T_{n}(q)$ on 
$End(({\mathbb C}^{2})^{\otimes  n})$
commutes with the action $\rho^n(\Delta_{}^n)$ of $ {\cal G}$, for
any $n$ \cite{PasquierSaleur90}.
Indeed \cite{Martin92}
\be
T_n(q) \cong \End_{U_q(sl(2))}(({\mathbb C}^{2})^{\otimes n})
\label{dual-pair}
\eeq
Applying this to ${\mathcal H}$ in (\ref{XXZ H}) we have the $U_q(sl(2))$
 symmetry of the Hamiltonian.

 For reference in the
{\em unfamiliar} generalization we are about to explore, let us
unpack the familiar duality (\ref{dual-pair}) a little.
For each choice of $q$, on
the one side we have a tower of algebras, and on the other a
single algebra, but equipped with coproduct, so that each has a
natural action on $V^{\otimes n}$ for each $n$. Each action lies in the
commutator of the other. In fact each action {\em is} the
commutator of the other \cite{Martin92}.

The questions raised are: What is the commutator of the $\Theta$
action of $T_n(q)$ on $V^{\otimes 2n}$? Can this commutator be constructed
in an analogously $n$-independent way? I.e., can it be constructed
   as the iterated coproduct of a bialgebra action? This would be
   the requirement for identifying the commutant as a quantum group
   symmetry of any associated spin chain. 
These are `big' questions. 
In section~\ref{rep thy} we answer the first one at least formally, by 
determining the structure of the commutator and showing that it is
   appropriately $n$-stable. 
Next, however, we answer affirmatively the slightly less ambitious question: 
Can any nontrivial part of the commutator of $b_n$ on $V^{\otimes 2n}$ be
constructed as an iterated coproduct?

\subsection{$\Uqsl2$ actions on $V \otimes V$}

By the general theory above, one may begin 
the search for ${\cal G}$-symmetries of ${\cal H} = {\cal H}^{\Theta}$ 
by looking at the ways in
which $V \otimes V$ can be equipped with the property of ${\cal G}$-module. 
We will think, in particular, of $\sigma({\cal G})$ acting on 
$V_{\lef 1} \otimes V_{\ri 1}$, and  $\sigma^{\otimes 2}(\Delta({\cal G}))$
acting on
$(V_{\lef 1} \otimes V_{\ri 1})\otimes(V_{\lef 2} \otimes V_{\ri 2})$,
and so on. 
\footnote{NB, These details are arbitrary for ${\cal G}$ itself, but we will be 
composing with a specific action of $b_n$, so it is necessary to be 
specific on this side also.
} 
Here we report on $\Uqsl2$ symmetries we have found 
(although we also show that this does not rule out other quantum groups). 
In terms of equivalence classes of actions we can essentially characterize any
such module by its irreducible content; and as already mentioned the
$\Uqsl2$ irreducibles are indexed by the whole numbers.
Since $V \otimes V$ is a four dimensional space, 
we have classes of modules of type 1+1+1+1, 1+1+2, 2+2, 1+3, 4. 
The first of these is 
$\epsilon\oplus\epsilon\oplus\epsilon\oplus\epsilon$, 
which is trivial.
For the others, of course, one is looking for commutation with a {\em
  specific} action of $b_n$, so one must check commutation for specific actions.
Since there are infinitely many of these in each case
a certain amount of sensible guesswork is required. 
Further, it is not clear that the parameter $q$ in $\Uqsl2$ need be
the same as the parameter $q$ in $b_n$, 
and we will not make this assumption!

For example, restricting to the case of {\em trivial} boundaries, 
one notes that 
the `real' and `mirror' sides of the construction of $\Theta$
are, in a suitable sense, decoupled. 
Thus there is an obvious pair of actions in the
$\rho\oplus\rho$
equivalence class, as follows.  

\subsection{A vestigial symmetry of the asymmetric chain 
            with $K^- \propto \Id$}

Consider representations of $U_{r}(sl(2))$ and $U_{\rs}(sl(2))$ respectively on
$V_{\lef 1} \otimes V_{\ri 1} = \mathbb C^{2} \otimes \mathbb C^{2}$:
\be
\rho_{1}(-) = \rho_{q=r}(w(-)) \otimes \Id , 
\qquad~~
\rho_{2}(-) = \Id \otimes \rho_{q=\rs}(-)
\label{sigma}
\eeq
NB, these are not tensor products of representations. 
Rather, regarded as an action on
$\mathbb C^{4} = \mathbb C^{2}\otimes \mathbb C^{2}$, 
each is simply isomorphic to $\rho \oplus \rho$
(for the appropriate choice of $q$-parameter).
Note that $\rho_{1}$ acts non-trivially on the `real' (left) space only,
whereas $\rho_{2}$ acts non-trivially on the mirror (right) space.
One is then to understand in the representation 
$\rho_1^{(2)}(-)$ on $(\C^2 \times \C^2)^{\otimes 2}$ given by 
\[
\rho_1^{(n)}(-) = \rho_{r}^{\otimes n}(\Delta^n(w(-))) \otimes \Id^n
\qquad
\rho_2^{(n)}(-) =  \Id^n \otimes \rho_{\rs}^{\otimes n}(\Delta^n(-))
\]
that the non-trivial factor acts on $V_{\lef 2} \otimes V_{\lef 1}$ 
and the trivial one on
$V_{\ri 1}\otimes V_{\ri 2}$ (and complementarily for $\rho_2$).
So by  
(\ref{comre1}) (or otherwise trivially)
\be
\Big [ \rho_{l}^{(2)} (x),\ (U(r^{-1}))_{\lef 1 \; \lef{2}} \Big ]
=0, ~~ \qquad
\Big [ \rho_{l}^{(2)} (x),\ 
              (U(\hat r))_{\ri 1 \; \ri{2}} \Big ] =0,
\eeq
for $l \in \{1,2\}$, $x\in {\cal G}$, 
so by (\ref{the})
\be
\Big [\rho_{l}^{(2)} (x),\ 
        \Theta( \abU_{ 1}) \Big ] =0, \qquad~~ l \in \{1,\ 2\}
\label{1a}
\eeq
By Proposition~\ref{commute n}, relations (\ref{1a}) imply 
corresponding statements for any number of sites $n$.
Recalling also (\ref{ht}) it follows (at least for $K^- = \Id$) that
\be
\Big[ \rho_{l}^{(n)}(x),\ {\cal H} \Big] = 0 . 
\label{12a} 
\eeq

Since the two quantum group actions commute with {\em each other} trivially,
we conclude that the model for $K^{-} = \Id$
has a $U_{r}(sl(2)) \otimes U_{\rs}(sl(2))$ symmetry.


This exercise serves to
illustrate the point that the close natural relationships between
spin-chain symmetry
and quantum group, which  we take for granted in conventional spin
chains, is not necessarily obvious in general.

The $K$-matrices (\ref{K3}) do not commute with the 
$\rho_l$ actions (\ref{sigma}), so
a nontrivial boundary term in the Hamiltonian 
breaks
the $U_{r}(sl(2)) \otimes U_{\rs}(sl(2))$ symmetry.

\subsection{Boundary stable symmetries}\label{Boundary stable}
\newcommand{\tq}{{\tilde q}}


Let $e_{ij}$ be the $4 \times 4$ elementary matrix:
$
(e_{ij})_{kl} = \delta_{ik}\ \delta_{jl}
$
and
\be
&&\tilde S^{+} = e_{14} = \sigma^+ \otimes \sigma^+
\qquad  \tilde S^{-} = e_{41} = \sigma^- \otimes \sigma^-
\qquad  \tilde S^{z} = {1\over 2}(e_{11} - e_{44}) 
\non\\  
&& S^{+} = e_{23} = \sigma^+ \otimes \sigma^-
\qquad S^{-} = e_{32}
\qquad  S^{z} = {1\over 2}(e_{22} - e_{33})
\eeq

For any $\tq$, there is a representation
$\sigma: U_{\tq}(sl(2)) \to \mbox{End}(\mathbb C^{2} \otimes \mathbb C^{2})$,
given by
\be
\sigma({\cal K})=  \tq^{-S^{z}},
\qquad~~ \sigma({\cal F})=S^{+} ,
\qquad~~ \sigma({\cal E})=S^{-}  
\label{gen3}
\eeq
(Note that $\sigma \cong \epsilon \oplus \rho \oplus \epsilon $,
but that it acts non-trivially on both the `real' and `mirror' spaces.)
In order to apply proposition~\ref{commute n} to commutation with $\Theta$ we
need to express $\Theta({\abU_i})$ in the form $M_{i \; i\!+\!1}$.
Our convention for the explicit action of $\sigma^{\otimes n}(\Delta^n)$ 
was chosen to make this possible. 
The explicit matrix form of  $\Theta({\abU_i})$ is the permutation of 
(\ref{rank 1 1}) given in \ref{A}. 
Direct calculation at $n=2$ then shows that 
$
\Big [\sigma^{\otimes 2}(\Delta({\cal K})),
\ \Theta(\abU_{1})\Big ] =0, 
$
and 
\be
  \Big [\sigma^{\otimes 2}(\Delta({\cal F})),
\ \Theta(\abU_{1})\Big ]_{kl} = 
-  \Big [\sigma^{\otimes 2}(\Delta({\cal E})),
\ \Theta(\abU_{1})\Big ]_{lk} = 
\hspace{2in}
\non\\ 
  \tilde q^{-{1\over 2}}(\tilde q -q)
\Big (\delta_{k6}(-r^{-1}\delta_{l4} - \hat r^{-1} \delta_{l13} +
\delta_{l10} -q^{-1}\delta_{l7}) - \delta_{l11}(-q^{-1}\delta_{k7} - r^{-1}
\delta_{k4} - \hat r^{-1} \delta_{k13} + \delta_{k10})\Big )
   \label{qq1} 
\eeq
(here we use the flattened labels $k,l \in \{ 1,2 , \ldots, 16 \}$). 
In summary  
\be
\Big[ \sigma^{\otimes 2}( \Delta(x)),\ \Theta(\abU_{1}) \Big] =0, 
~~~\forall x \in \Gq{\tilde q} 
~~~\iff \tilde q =q
\label{qg3} 
\eeq
Direct calculation at $n=1$ shows  (for any ${\tilde q}$) that 
\be
\Big[ \sigma(x),\ \Theta^{ii}(\abe) \Big]  
= \Big [\sigma(x),\ \KI{ii}(\lambda) \Big]  =0 
\qquad \forall x \in \Gq{\tilde q}
\label{cr01b} 
\eeq
but that there is no such commutation for $\Theta^i$. 
By Proposition~\ref{commute n}, relations (\ref{qg3}), (\ref{cr01b}) imply 
\prl(symm q)
For every $n$, 
$ 
\Big[ \Theta(\abU_i) ,\  \sigma^{\otimes n} (\Delta^{n}(x)) \Big] =
\Big[ \Theta^{ii}(\abe) ,\  \sigma^{\otimes n} (\Delta^{n}(x)) \Big] 
=0 , 
$
and hence the Hamiltonian ${\cal H} = {\cal H}^{\Theta^I}$ (\ref{ht}) 
with boundary term of type ($ii$) or (0) 
has a $U_{q}(sl(2))$ symmetry:
\be
   \Big [{\cal H} ,\  \sigma^{\otimes n} (\Delta^{n}(x)) \Big ] = 0 . 
\label{2b}
\label{comhkb} 
\eeq
\end{pr}

Another representation
$\pi: U_{\tilde q}(sl(2)) \to \mbox{End}( \C^{2} \otimes \C^{2})$ is
\be
\pi({\cal K}) =\tq^{\tilde S^{z}},
\qquad~~ \pi({\cal E})=\tilde S^{+},
\qquad~~ \pi({\cal F})=\tilde S^{-}. 
\label{gen1}
\eeq
\newcommand{\ii}{\sqrt{ \! - \! 1 }}%
One can show directly that
$
\Big [\pi^{\otimes 2}(\Delta({\cal K})),\ \Theta(\abU_{1})\Big ] =0,
$
and
\be
&& \Big [\pi^{\otimes 2}(\Delta({\cal F})),\ \Theta(\abU_{1})\Big]_{kl} =
 - \Big [\pi^{\otimes 2}(\Delta({\cal E})),\ \Theta(\abU_{1})\Big]_{lk} =
\non\\ 
&&  \tilde q^{-{1\over 2}}(\tilde q -\ii) \Big( \delta_{k16}(\delta_{l4} -
\hat r^{-1} \delta_{l7} - r \delta_{l10} +i\delta_{l13}) -
\delta_{l1}(\delta_{k4} - \hat r^{-1} \delta_{k7} - r \delta_{k10}
+i\delta_{k13})\Big ) 
\label{qi1} 
\eeq 
In summary  
\be
\Big [\pi^{\otimes 2}( \Delta(x)),\ \Theta(\abU_{1}) \Big ] = 0
\qquad \forall x \in \Gq{\tilde q} 
\qquad \iff \tq = \ii
\label{2a}
\eeq
Direct calculation shows (see (\ref{K3}) and (\ref{gen1})) that
\be 
\Big [ \pi(x),\ \Theta^i(\abe) \Big]  
= \Big[\pi(x),\ \KI{i}(\lambda) \Big] = 0
\qquad \forall x \in \Gq{\tilde q}
\label{cr01}. 
\eeq  
From (\ref{2a}) and (\ref{cr01}) and Proposition~7 we have
\prl(symm q pi)
For every $n$, the Hamiltonian ${\cal H} = {\cal H}^{\Theta^I}$ (\ref{ht}) 
with boundary term of type ($i$) or (0)  
has a $U_{\ii}(sl(2))$ symmetry: 
\be 
\Big[ {\cal H},\ \pi^{\otimes n}(\Delta^{n}(x)) \Big] =0. 
\label{comhk} 
\label{2aa}
\eeq
\end{pr}

\newcommand{\lambent}{exposed}



In summary: 
we have exposed a 
$U_{q}(sl(2)) \otimes U_{i}(sl(2))$ symmetry of the
trivial boundary Hamiltonian.
The $U_{q}(sl(2))$ symmetry is preserved in the
presence of the nontrivial boundary ($ii$), 
but broken by the boundary ($i$);
whereas the $U_{i}(sl(2))$ symmetry is
preserved in the presence of ($i$), but broken by  ($ii$). 
(For comparison, the Hamiltonians 
considered here are written out in terms of Pauli matrices in
\ref{D}.)  


While it is possible that the symmetry descibed here
constitutes the full ${\cal G}$-symmetry of
${\cal H}$ in each boundary type,
it is {\em not} the full commutator as it is in the XXZ case.
To see this is a managable exercise, comparing the generic simple
multiplicities in $\Theta$ 
(given in \cite{MartinRyom03} in some cases) with the corresponding
data for the quantum group, as we will now show. 
\\
Non-trivial boundaries such as ours sometimes allow symmetries 
called boundary quantum
algebras \cite{MezincescuNepomechie97,DeliusMackay03,Doikou04a,Doikou04b}. 
These will be discussed in  Section~\ref{bqa}.


\subsection{Representation theory of $\End_{b_n}(V^{\otimes 2n})$} \label{rep thy}
\newcommand{\ite}{\it}
By (\ref{ht}) the spectrum of ${\cal H}$ is determined (in principle) by
the representation theory of $T_n$ and $b_n$. In practice this theory
is not yet an effective substitute for Bethe ansatz
(cf. \cite{MartinSaleur94c}), but some useful
results may be gleaned by using the approaches in tandem. 
For example,  Martin and Saleur's extension of the representation $\rrep{q}$ to
$b_n$ \cite{MartinSaleur94a} defines the spin-1/2 XXZ chain with 
non-diagonal boundaries.%
\footnote{Apart from the significant disadvantage of not having
reference states, this is the obvious candidate for studying
non-trivial boundaries, and has received significant attention 
--- see \cite{Nepomechie03} for references.}
By our general theory, therefore, this chain  has the same spectrum, up to 
multiplicities, as the \gemini\ chain, if and only if the underlying
representations have the same irreducible content up to 
multiplicities.
We may derive the irreducible content of the $\Theta$ representations
(and other tensor space representations such as $\rrep{q}$) as follows. 

\newcommand{\lmod}{\!-\!\mod}%

Let $A,B$ be algebras. An   idempotent $e \in A$ 
is called a {\em localisation  idempotent} from $A$ to $B$ if
$eAe \cong B$. Through this isomorphism one may regard $eA$ as a left
$B$ right $A$ bimodule. 
Thus we have a functor 
$L:A\lmod \rightarrow B\lmod$
given by $L(M) = eM$
(here $A\lmod$ denotes the category of left $A$-modules);
and a functor 
$G:B\lmod \rightarrow A\lmod$
given, via the isomorphism, by $G(N) = Ae \otimes_B N$.  
Thus, presence of a localisation idempotent ensures that there is a full
embedding of the category of left (or right) $B$-modules in the
corresponding category of $A$-modules
(the functor $L$ is called localisation, and is exact; 
and $G$, globalisation, is right exact). 
For example, appropriately normalised, $\abU_{n-1}$ is a localisation
idempotent from $T_n$ to $T_{n-2}$, and also from $b_n$ to $b_{n-2}$. 
Both of these collections of algebras have {\em standard modules}
(generically irreducible modules) $ \Delta_{\nu}^{n}$, 
with the following properties. 
\prl(globloc)
The standard modules of $T_n$ are indexed by 
$\nu \in \{ n, n-2, \ldots, 1/0 \}$. 
The standard modules of $b_n$ are indexed by 
$\lambda \in \{ -n, -(n-2), \ldots, (n-2), n \}$. Then 
\\ 
(i) localisation takes a standard module to a standard module
$ \Delta_{\nu}^{n}    \mapsto \Delta_{\nu}^{n-2}$, 
or zero if $| \lambda |  > n-2$. 
\\
(ii) localisation takes $\Theta:b_n$ to $\Theta:b_{n-2}$ (any $I$).
\\
(ii') localisation takes $\rrep{q}:b_n$ to $\rrep{q}:b_{n-2}$.
\\
(iii) restriction of standard modules is given by
$ \Delta_{\nu}^{n} 
     \mapsto \Delta_{\nu+1}^{n-1}+\Delta_{\nu-1}^{n-1}$,
\\
(iv) restriction takes 
$\Theta:b_n \mapsto 4(\Theta:b_{n-1})$ (any $I$). 
\\
(iv') restriction takes 
$\rrep{q}:b_n \mapsto 2(\rrep{q}:b_{n-1})$. 
\end{pr}
{\em Proof:}
The index theorems and 
{\ite (i)} are standard results \cite{MartinRyom03};
{\ite (ii)}  follows from equation(\ref{the rank 1 property});
{\ite (iii)} is again well known; 
{\ite (iv)} follows from a simple calculation. 
\Qed


For $M$ a module with a well defined filtration by standard modules
(such as $\Theta$), 
let us write $(M:\Delta_{\nu})$ for the multiplicity of
$\Delta_{\nu}$ in $M$. 
By {\ite (i)} and {\ite (ii)} we have 
\begin{co}\label{co10.1}
The multiplicity   
$(\Theta : \Delta_{\nu}^{n})$
does not depend on $n$, once $n \geq | \nu|$. 
(Similarly for $\rrep{q}$.)
\end{co}
By \ref{co10.1}, {\ite (iii)} and {\ite (iv)} we have
\begin{co}
Let $\C^{\infty}$ have basis $\{ e^{\nu} \; : \; \nu \in \Z \}$. 
Setting 
$\C^{\infty} \ni v = \sum_{\nu} (\Theta : \Delta_{\nu}^{n}) e^{\nu}$
and operator $\Chi e^{\nu} =  e^{\nu +1} + e^{\nu -1}$
then $\Chi v = 4v$. 
\end{co} 
Given the multiplicities for $\nu < n$, it follows that
the only question at level $n$ is the mutliplicities of 
the one-dimensional modules $\Delta_{\pm  n}^n$. 
The combined multiplicity is fixed by the
dimension of $\Theta$ (for example in the $b_n$ case):
$$
(\Theta : \Delta_{n}^{n})+(\Theta : \Delta_{-n}^{n})
= \dim(\Theta) 
- \sum_{|\nu| <n} \dim(\Delta_{\nu}^{n}) (\Theta : \Delta_{\nu}^{n}) ;
$$
and the individual multiplicities by (iii) and (iv). 

It follows that the multiplicities for 
{\em all} $n$ are determined by the case $n=1$
and either the case $n=2$ or $n=0$ (if this can be appropriately
defined --- in our case it denotes a copy of the ground ring). 
This proves proposition~\ref{big one}. 
The formal case $n=0$ does not depend on the extension to $\Theta$. 
By construction $\Theta^{i}$ and $\Theta^{ii}$ have the same
multiplicities at $n=1$, so we deduce that they are generically
isomorphic. Thus the corresponding spin chains will have the same
spectrum. (As we will see later, this abstract proof does not
correspond to a straightforward connection at the level of Bethe
Ansatz.)

Since the dimensions of the standard modules, and of $\Theta$, 
are readily computed,
the spectrum may be determined explicitly. 

Firstly for the $T_n$ case:
Starting with the
multiplicity 1 for $\Delta_{0}^0$ (and hence for $\Delta_{0}^{n}$ for
any even $n$), and the multiplicity 4 for $\Delta_{1}^1$  
(and hence for $\Delta_{1}^{n}$ for any odd $n$)
we deduce the
multiplicity 15 for $\Delta_{2}^2$ in the $4^n=16$ dimensional
representation $\Theta$ in case $n=2$,
and then $(\Theta:\Delta_{3}^3) = 4^3 - 2\times4$. 
One way to proceed is to  tabulate the dimensions of the standard
modules in each layer explicitly, 
as  in the lower part of the following layout. 
The upper part then gives the multiplicities, 
which we have entered as the vertices of a graph.
The edges of this graph serve to indicate the restriction rules. 
It follows that the sum of the multiplicities on the vertices adjacent
to a given vertex is 4 times that on the vertex itself.
\[
\xymatrix@R=15pt@C=12pt@M=1pt{
\nu: & 0 & 1 & 2 & 3 & 4 &   \\
(\Theta:\Delta_{\nu}^{odd}): 
         &   & 4 \ar@{-}[dr]^{} &   & 56 \ar@{-}[dr]^{} &&780& \\
(\Theta:\Delta_{\nu}^{even}): 
         & 1 \ar@{-}[ur]^{} &   & 15 \ar@{-}[ur]^{} &   &209\ar@{--}[ur]  \\
\hspace{-.95in} \dim(\Delta^n_{\nu}) \qquad 
n=1               &   & 1         \\
n=2               & 1 &   &  1    \\
n=3               &   & 2 &   & 1 \\
n=4               & 2 &   & 3 &   & 1    
}
\]
And so on.%
\footnote{The multiplicity combinatorics may be encoded in other ways too.
Consider
\[
\underbrace{\mat{cccccccccccc} 
0 & 1 \\
1&0&1 \\
&1&0&1 \\
&&1&0&1 \\
&&&&\ddots
\tam}_{\Chi^+} 
\mat{c} 1 \\ 4 \\ 15 \\ 56 \\ \vdots 
\tam 
= 4 
\mat{c} 1 \\ 4 \\ 15 \\ 56 \\ \vdots 
\tam 
\]
See \cite{MartinRyom03} for further details.
}

In case $\Theta^i$  (or $\Theta^{ii}$) the corresponding picture is:
\[
\xymatrix@R=15pt@C=12pt@M=1pt{
\nu:& -3&-2&-1& 0 & 1 & 2 & 3 & 4   \\
(\Theta^I:\Delta_{\nu}^{odd}): 
        & 11 \ar@{-}[dr]^{}  & & 1 \ar@{-}[dr]^{} &  & 3 \ar@{-}[dr]^{} &   & 41
\ar@{--}[dr]^{} &     \\
(\Theta^I:\Delta_{\nu}^{even}): 
        & &3 \ar@{-}[ur]^{} && 1 \ar@{-}[ur]^{} &   & 11 \ar@{-}[ur]^{} &   &153  \\
\hspace{-.95in} \dim(\Delta^n_{\nu}) \qquad 
n=0     & &&   & 1         \\
n=1     & && 1 &   &  1    \\
n=2     & &1&   & 2 &   & 1 \\
        &1& & 3 &   & 3 &   & 1    
}
\]
And so on. 

In case $\Theta^+$ (resp. $\rrep{q}$ of \cite{MartinSaleur94a}) 
the corresponding picture is:
\[
\xymatrix@R=15pt@C=12pt@M=1pt{
\nu:& -3&-2&-1& 0 & 1 & 2 & 3 & 4   \\
(\Theta^+:\Delta_{\nu}^{odd}): 
        & 26 \ar@{-}[dr]^{}  & & 2 \ar@{-}[dr]^{} &  & 2 \ar@{-}[dr]^{} &   & 26
\ar@{--}[dr]^{} &     \\
(\Theta^+:\Delta_{\nu}^{even}): 
        & &7 \ar@{-}[ur]^{} && 1 \ar@{-}[ur]^{} &   & 7 \ar@{-}[ur]^{} &   &97  \\
(\rrep{q}:\Delta_{\nu}^{odd}): 
        & 1 \ar@{-}[dr]^{}  & & 1 \ar@{-}[dr]^{} &  & 1 \ar@{-}[dr]^{} &   & 1
\ar@{--}[dr]^{} &     \\
(\rrep{q}:\Delta_{\nu}^{even}): 
        & &1 \ar@{-}[ur]^{} && 1 \ar@{-}[ur]^{} &   & 1 \ar@{-}[ur]^{} &   &1  \\
}
\]
which shows that there is  not a unique groundstate for odd $n$ in
the $\Theta^+$  case. 

Comparing the multiplicities in the $\Theta$ cases 
with the actions of the quantum groups
described above we see that the naive commutant is much bigger than
the set of commuting matrices given by these actions.
In section~\ref{bqa} we discuss  
quantum group 
actions which 
do not lie in the commutant themselves, but 
will
underly the exposition of further symmetries, 
and also help us prepare to address
the technical question of Bethe Ansatz asymptotics. 
On the other hand, since all multiplicities here are nonzero we have
proved 
\prl(spectrum_upto)
The spectrum of \gemini\ and boundary XXZ are the same up to multiplicities.
\end{pr}

   


\section{Further symmetries} \label{bqa}

In this section we investigate further the symmetry of the open
Hamiltonian ${\cal H}^{\Theta}$  
in the presence of non-trivial boundaries
(\ref{K3}). We will first need some notation. 

\subsection{The quantum affine   
   algebra $\AUqsl2$}
The Cartan matrix of the affine Lie algebra ${\widehat {sl(2)}}$
\cite{Kac90} is 
\be
   (a_{ij}) = \left( \begin{array}{cc}
   2   &-2\\
   -2  &2\\
   \end{array} \right)
   \,
\eeq
The quantum Kac--Moody  algebra $\AUqsl2 $
has Chevalley-Serre generators
\cite{Jimbo85a,Jimbo85b,Drinfeld86}  
$e_{i}$, $f_{i}$, $k_{i}$, $i=1,2$ with defining relations
\be
   k_{i}\ k_{j} = k_{j}\ k_{i},
   ~~ \qquad k_{i}\ e_{j}&=&q^{{1\over 2}a_{ij}}e_{j}\ k_{i},
   ~~ \qquad k_{i}\ f_{j}
   = q^{-{1\over 2}a_{ij}}f_{j}\ k_{i}, \non
   \\
   \Big [e_{i},\ f_{j}\Big ] &=& \delta_{ij}{k_{i}^{2}-k_{i}^{-2} \over q-q^{-1}},
\label{aff comm}
\eeq
   (and the $q$ deformed Serre relations)
\be
e_{i}^{3}\ e_{j} -[3]_{q}\ e_{i}^{2}\ e_{j}\ e_{i} +[3]_{q}\
e_{i}\ e_{j}\ e_{i}^{2} -e_{j}\ e_{i}^{3} &=&0 \non
\\
f_{i}^{3}\ f_{j}
-[3]_{q}\ f_{i}^{2}\ f_{j}\ f_{i}+[3]_{q}\ f_{i}\ f_{j}\ f_{i}^{2}
-f_{j}\ f_{i}^{3} &=& 0,
~~ \qquad i\neq j.
\label{serre}
\eeq

Note that if $\phi$ is a representation of $U_{q}(sl(2))$ then
the map $\phi_0$ on generators given  by
$\phi_0(e_1) = \phi(\cE)  \;$,
$\phi_0(f_1) = \phi(\cF)  \;$,
$\phi_0(k_1) = \phi(\cK)  \;$,
$\phi_0(e_2) = \phi(w(\cE))$,
$\phi_0(f_2) = \phi(w(\cF))$,
$\phi_0(k_2) = \phi(w(\cK))$
satisfies $(\ref{aff comm})$.
In particular if  $\phi(\cE^2)=0$ 
(i.e. $\phi$ is a sum of representations of class $\epsilon$ and $\rho$)
then
$\phi_0$ extends to a representation of $\AUqsl2$.

Note also that the relations (\ref{aff comm}), (\ref{serre}) are, except for the
$i=j$ case of the commutator, homogeneous in each generator.
It follows that given a representation, there is another with
$\psi(e_i) \rightarrow c \psi(e_i)$,
$\psi(f_i) \rightarrow c^{-1} \psi(f_i)$,
for either $i$ and any constant $c \neq 0$.
In particular,
for $\lambda$ a scalar, there is a representation
$\rho_{\lambda}: \AUqsl2 \to End(\mathbb C^{2})$
given by
\be
\rho_{\lambda}(e_{1}) &=& \sigma^{\pm}, ~~ \qquad
\rho_{\lambda}(f_{1}) = \sigma^{\mp}, ~~ \qquad
\rho_{\lambda}(k_{1}) = q^{\pm\sigma^{z}} \non
\\
\rho_{\lambda}(e_{2}) &=& e^{-2\mu \lambda} \sigma^{\mp}, ~~ \qquad
\rho_{\lambda}(f_{2}) =e^{2\mu \lambda} \sigma^{\pm},~~ \qquad
\rho_{\lambda}(k_{2})= q^{\mp \sigma^{z}}.
\label{action}
\eeq
This is  the {\it evaluation representation} \cite{Jimbo85a}.
If $\phi$ above obeys $\phi(\cE^2)=0$ we call $\phi$ {\em linear}.
If $\phi$ is linear the corresponding construction 
$\phi \rightarrow \phi_{\lambda}$ gives a representation of
$\AUqsl2$.

Set ${\cal A} = \AUqsl2$.
A Hopf algebra structure is defined on ${\cal A}$ by introducing the coproduct%
\footnote{Also define the co-unit $\epsilon$ and the antipode $S$:
\be
\epsilon(e_{i})=\epsilon(f_{i}) =0, 
~~\epsilon(k_{i}^{\pm 1}) =1 , 
\qquad
S(e_{i}) = -q^{-1}e_{i}, 
~~S(f_{i}) =-qf_{i}, 
~~S(k_{i}^{\pm 1}) = k_{i}^{\mp 1}. 
\non
\eeq
}
$\Delta: {\cal A}\ \to {\cal A} \otimes {\cal A}$:
\be
\Delta(e_{i}) &=& k_{i}^{-1} \otimes e_{i} + e_{i} \otimes k_{i} \non\\
\Delta(f_{i}) &=& k_{i}^{-1} \otimes f_{i} + f_{i} \otimes k_{i} \non\\
 \Delta(k_{i}^{\pm 1}) &=& k_{i}^{\pm 1} \otimes k_{i}^{\pm 1}.
\eeq
\subsection{Boundary quantum algebra symmetry}

We now construct, cf. \cite{DeliusMackay03,Doikou04a,Doikou04b}, 
realizations of the so called boundary quantum algebra
   generators which commute with ${\cal H}^{\Theta^{I}}$.
We shall need versions of the evalutation representation (\ref{action}) 
applied to (\ref{sigma}), (\ref{gen1}), (\ref{gen3}):
$\pi \to \pi_{\lambda}, 
~\sigma \to \sigma_{\lambda}, 
~\rho_{1} \to \rho^{1}_{\lambda}, 
~\rho_{2} \to \rho^{2}_{\lambda}$.
 \noindent
E.g. 
$\sigma_{\lambda}: U_{q}(\widehat{sl(2)})\to \mbox{End}(\mathbb C^{4})$:
\be 
\sigma_{\lambda}(k_{1}) &=&q^{-S^{z}},
~~\sigma_{\lambda}(e_{1})=S^{-},
~~~\sigma_{\lambda}(f_{1})=S^{+} 
\non\\
\sigma_{\lambda}(k_{2})
&=&q^{S^{z}},
~~\sigma_{\lambda}(e_{2})=e^{-2\mu \lambda} S^{+},
~~~\sigma_{\lambda}(f_{2})=e^{ 2\mu \lambda}S^{-}
\label{action2} 
\eeq


  \noindent
 Similarly, let
  $\pi_{\lambda}: U_{i}(\widehat{sl(2)})\to \mbox{End}(\mathbb C^{4})$
  such that
\be
\pi_{\lambda}(k_{1}) &=& i^{\tilde S^{z}},
~~~\pi_{\lambda}(e_{1})= \tilde S^{+},
~~~\pi_{\lambda}(f_{1})=\tilde S^{-}  \non
\\
\pi_{\lambda}(k_{2}) &=& i^{-\tilde S^{z}},
~~~\pi_{\lambda}(e_{2})= e^{-2\mu \lambda}\tilde S^{-},
~~~\pi_{\lambda}(f_{2})=e^{ 2\mu \lambda}\tilde S^{+}, 
\label{action3} 
\eeq
where $\mu$ is derived from the {\em blob algebra} parameter $q$, 
not from $\tilde q = i$.

   \newcommand{\appa}{\alpha}
   
For $z\in \{q, i, r, \hat r \}$, 
consider the element of $U_{z}(\widehat{sl(2)})$ given by  
\be
{\cal Q}_{j}^{z} &=&
   z^{-{1\over 2}}k_{j}e_{j}
+  z^{ {1\over 2}}k_{j}f_{j} 
+  x_{z}^{j} k_{j}^{2}   -  x_{z}^{j}I, 
\qquad  j \in \{ 1,\ 2 \}
\label{bg}
\eeq
(the scalars $x_{z}^{j}$ will be identified later). 
Define  
${\cal B} \; = \; 
< \! Q_{1}^{z},\ Q_{2}^{z} \! > \; \subset \; {\cal A}\vert_{q=z}$.
These elements have a relatively simple expression for the 
iterated coproduct inherited from ${\cal A}$, i.e.
\be
   \Delta^{n}({\cal Q}_{i}^{z})
   = I \otimes \Delta^{(n-1)}({\cal Q}_{i}^{z}) + {\cal Q}_{i}^{z} \otimes
   \Delta^{(n-1)}(k_{i}^{2}).
\label{coba}
\eeq
   Note however that this is not closed on ${\cal B}^{\otimes n} $.


Let us now consider various actions of ${\cal B}$ 
on $V^{\otimes 2n}$ (by restriction of the
   ${\cal A}$ action), and hence on
   the solutions (\ref{K3}i) and (\ref{K3b}ii)
   and the corresponding Hamiltonians ${\cal H}^{\Theta^{I}}$.
   
\vskip 0.1cm

\noindent {\bf Type (i)}
Guided by the XXZ case \cite{Doikou04a,Doikou04b} 
we make the following identifications for the solution ($i$): 
\be 
\label{doink}
   x_{q}^{1} = {Q -Q^{-1}\over q-q^{-1}}, 
~~ \qquad 
x_{q}^{2} =  {Q+Q^{-1} \over 2
\delta_{e} \sinh^{2} i\mu}\cosh 2 i
   \mu \zeta.
\label{xi1}
\eeq
   It can be shown by direct calculation that
\be
\sigma_{\lambda} ({\cal Q}_{j}^{q})\ \KI{i}(\lambda)
  = \KI{i}(\lambda)\  \sigma_{-\lambda} ({\cal Q}_{j}^{q})
\qquad j \in \{1,2 \}
\label{cr1} 
\eeq
   provided (\ref{doink}) holds.
From (\ref{cr1}) it can be shown that for $n=1$
\be
   \Big [\sigma_{\lambda} ({\cal Q}_1^{q}),\ \Theta^{i}(\abe)\Big] =0, 
\label{inter0}
\eeq
Taking into account
   the commutation relations (\ref{qg3}) and (\ref{bg}), (\ref{inter0}) and
Proposition~\ref{commute n} then
\be 
\Big[ \sigma_{\lambda}^{\otimes n}
   (\Delta^{n}({\cal Q}_{1}^{q})),\ \Theta({\cal U}_{l}) \Big]
= \Big[ \sigma_{\lambda}^{\otimes n} (\Delta^{n}({\cal Q}_{1}^{q})),\
   \Theta^{i}(\abe) \Big]=0
\eeq
(any $l$) so  
\be
\Big[ \sigma_{\lambda}^{\otimes n} (\Delta^{n}({\cal Q}_{1}^{q})),
\ {\cal H}^{\Theta^{i}} \Big]=0. 
\label{comh}
\eeq
We conclude from (\ref{comh}) that the presence of the  boundary
   (\ref{K3}i) does not break the $U_{i}(sl(2))$ part of the \lambent\
   symmetry (\ref{comhk}), and also preserves the `charge' 
$\sigma_{\lambda}^{\otimes n} (\Delta^{n}({\cal Q}_{1}^{q}))$  (\ref{comh}).

   \vskip 0.1cm

\noindent {\bf Type (ii)} 
Considering the solution ($ii$) we make the identifications:
\be 
x_{i}^{1} = - {Q-Q^{-1}\over 2i}, 
~~ \qquad 
x_{i}^{2} =  {Q+Q^{-1} \over  2i
\delta_{e} \sinh i\mu}\cosh 2 i \mu \zeta, 
\label{xi2} 
\eeq
where again $\mu$ is the blob algebra parameter. 
Then it can be shown that
\be 
\pi_{\lambda}({\cal Q}_{1}^{i})\ \KI{ii}(\lambda) =
   \KI{ii}(\lambda)\ \pi_{-\lambda}({\cal Q}_{1}^{i}) 
\label{cr2}
\eeq
   whereas
\be
\pi_{\lambda}({\cal Q}_{2}^{i})\ \KI{ii}(\lambda) 
= \KI{ii}(\lambda)\  \pi_{-\lambda}({\cal Q}_{2}^{i}) 
~~
\iff ~Q = e^{-{i\mu \over 2} -{i\pi \over 4}}.
\label{cr2b}
\eeq 
From (\ref{cr2}) it follows
\be
   \Big [\pi_{\lambda} ({\cal Q}_{1}^{i}),\ \Theta^{ii}(\abe) \Big] =0.
\label{inter1}
\eeq 
Taking into
   account the commutation relations (\ref{2a}) and (\ref{bg}), (\ref{inter1}), and
   Proposition~\ref{commute n} it can be directly shown that
\be
\Big [\pi_{\lambda}^{\otimes n}
(\Delta^{n}({\cal Q}_{1}^{i})),\ \Theta({\cal U}_{i}) \Big]
= \Big[ \pi_{\lambda}^{\otimes n} (\Delta^{n}({\cal Q}_{1}^{i})),
\ \Theta^{ii}(\abe) \Big] = 0
\eeq
so 
\be
   \Big
   [\pi_{\lambda}^{\otimes n} (\Delta^{n}({\cal Q}_{1}^{i})),
\ {\cal H}^{\Theta^{ii}} \Big]=0. 
\label{comhb}
\eeq
Equation (\ref{comhb}) implies that the presence of the boundary (\ref{K3b}ii), 
which breaks the $U_{i}(sl(2))$ ($\pi$) part of the \lambent\
   symmetry (\ref{comhkb}), 
preserves the `charge' $\pi_{\lambda}^{\otimes n}
(\Delta^{n}({\cal Q}_{1}^{i}))$.  

Symmetries in the cases (+) and (iii) are relegated to a footnote
for brevity.
\footnote{{
\noindent   {\bf Type (+)} 
The presence of boundary type (+)
breaks both $U_{q}(sl(2))$ and $U_{i}(sl(2))$ symmetries.  
However it is clear from the form of the solution $\KI{+}$  and from
relations (\ref{cr01}), (\ref{cr01b}), (\ref{cr1}) and (\ref{cr2}) 
(which also hold for 
$\Theta^i(\abe)$ (\ref{cabl}) and $\Theta^{ii}(\abe)$ (\ref{cabl2})) that
\be 
&& \sigma_{\lambda}({\cal Q}_{1}^{q})\ \KI{+}(\lambda) 
=  \KI{+}(\lambda)\ \sigma_{\lambda}({\cal Q}_{1}^{q}) 
\non\\ 
&& \pi_{\lambda}({\cal Q}_{1}^{i})\ \KI{+}(\lambda) 
=  \KI{+}(\lambda)\ \pi_{\lambda}({\cal Q}_{1}^{i}).
\label{int+} 
\eeq 
provided that relations (\ref{xi1}), (\ref{xi2}) hold simultaneously. 
It is clear from (\ref{comh}), (\ref{comhb}),  (\ref{int+})  that
\be 
\Big [\sigma_{\lambda}^{\otimes n} (\Delta^{n}({\cal Q}_{1}^{q})),\ {\cal H}
\Big ] =0, ~~~\Big [\pi_{\lambda}^{\otimes n} (\Delta^{n}({\cal Q}_{1}^{i}),\ {\cal
H} \Big ] =0. 
\label{symm0} 
\eeq 

\noindent {\bf Type (iii)} 
In case (iii) both symmetries 
$U_{q}(sl(2))$, $U_{i}(sl(2))$ 
(and $U_{r}(sl(2))$, $U_{\hat r}(sl(2))$) are broken.
However one can explicitly show that
\be
&&\rho_{\lambda}^{1} ({\cal Q}_{1}^{r})\ \KI{iii}(\lambda)
= \KI{iii}(\lambda)\ \rho_{\lambda}^{1} ({\cal Q}_{1}^{r}) 
\non\\ 
&&\rho_{\lambda}^{2} ({\cal Q}_{1}^{\hat r})\ \KI{iii}(\lambda)
= \KI{iii}(\lambda)\ \rho_{\lambda}^{2} ({\cal Q}_{1}^{\hat r})
\label{cr1c} 
\eeq  
provided
\be  
x_{\hat r}^{1}= {\sqrt{iQ} -\sqrt{-iQ^{-1}}\over \hat r-\hat r^{-1}},
~~~x_{r}^{1} = i{\sqrt{iQ} +\sqrt{-iQ^{-1}}\over r-r^{-1}} \label{xi3}
\eeq 
and  $\rho_{i}$ is given by (\ref{sigma}). Again we obtain 
\be 
\Big[ {\cal H},\ \rho_{\lambda}^{1\otimes n}( \Delta^n({\cal
   Q}_{1}^{r})) \Big ] =0, 
~~ \Big[ {\cal H},\ \rho_{\lambda}^{2\otimes n}(
   \Delta^n({\cal Q}_{1}^{\hat r})) \Big ] =0. 
\label{symm} 
\eeq

It should be pointed out that we were able to derive intertwining
   relations (\ref{cr1}), (\ref{cr2}), (\ref{cr2b}) between the
   generators of ${\cal B}$  
and the $K$-matrices of type ($i$) and ($ii$), 
whereas for the other solutions we proved
commutation relations only for the non-affine generators.
}}
\section{Spectrum and Bethe ansatz equations}\label{bethe}
   Our  aim in what follows is to find the exact spectrum of the open
 \AC\ spin chain Hamiltonian,
   in types ($i$) and ($ii$), and study the role of the boundary parameter. Our
objective in
   the remainder of the present paper is to determine the {\em form} of
   the spectrum, and   to derive the  Bethe ansatz equations.
   For numerical solutions we need the large $\lambda$ asymptotics which,
   as already explained, will be treated elsewhere.

\newcommand{\yyp}{p}  

Reparameterizing by  $ Q = iq^p = ie^{i \mu \yyp}$
and setting
\be
\delta_{e} =-{\sinh i \mu \yyp \over \sinh (i \mu)}
\eeq
we have the following formulae 
from (\ref{kap}, \ref{kapp})
which will be useful later
\be \label{(p=m)}
\kappa^{i} = {\sinh i\mu (\yyp-1) \over \sinh (i \mu)} \label{rs}
\\
\kappa^{ii} = {\sinh i \mu ( \yyp +{\pi \over 2 \mu} )\over \sinh (i\mu)}.
\label{rs2}
\\
\kappa^{iii} = {\sinh i\mu(\yyp -1) -i \over \sinh (i \mu)}  \label{rs3}
\eeq
(The parameter $\yyp$ in type ($i$) can be identified with the usual $\yy$ of
 the parameterization $b_n(q,\yy)$
 which appears in the literature \cite{MartinRyom03}. 
 This parameterization is
 natural both from the point of view of representation theory, and
 integrability of the corresponding spin chain.
Note that $\yyp$ in types ($i$) and ($ii$) can be related by identifying
the ratios $\frac{\delta_e}{\kappa^{i}}$ and $\frac{\delta_e}{\kappa^{ii}}$, but
$\yyp$ in type ($ii$) does not have the same simple relation with the `usual'
parameterization $b_n(q,\yy)$. Nonetheless it is crucial for the
analysis of the `type ($ii$)' spin chain --- see later.)

\subsection{Reference states} \label{ref states}

\newcommand{\pseudovacuum}{pseudo-vacuum}
\newcommand{\pseudovacua}{pseudo-vacua}

Since the \gemini\ model conserves `charge' in a manner analogous to
trivial boundary XXZ, 
the first step 
is to find the standard ordered basis elements of $V^{\otimes 2n}$ which are
already eigenstates of the transfer matrix (\ref{transfer})
--- the \pseudovacua.
Define
\[
| 00 \rangle = {1 \choose 0} \otimes {1 \choose 0}
\qquad
| 01 \rangle = {1 \choose 0} \otimes {0 \choose 1}
\qquad
| 11 \rangle = {0 \choose 1} \otimes  {0 \choose 1}
\qquad
| 10 \rangle = {0 \choose 1} \otimes  {1 \choose 0} .
\]
\be \label{ref11}
|\omega_{+}\rangle = \underbrace{{1 \choose 0} \otimes \cdots
\otimes {1 \choose 0}}_{2n}
= \otimes_{i=1}^{n}  | 00 \rangle_{i} \,
\eeq
\be
|\omega_{2}\rangle = \otimes_{i=1}^{n} | 01 \rangle_{i} 
\label{ref11b} 
\qquad
|\omega_{1} \rangle=  \otimes_{i=1}^{n}| 11 \rangle_{i} 
\qquad
|\omega_{3}\rangle =  \otimes_{i=1}^{n}| 10 \rangle_{i} 
\eeq
The \pseudovacua\ 
$|\omega_{+}\rangle,|\omega_{2}\rangle,|\omega_{1}\rangle,|\omega_{3}\rangle$ 
are all eigenstates in the case $K^- \propto \Id$.

{\bf Type (i)} 
Consider solution (i) (\ref{K3}). 
The first reference state is the \pseudovacuum\ $|\omega_{+}\rangle$. 
The \pseudovacuum\ eigenvalue takes the form
\be
\Lambda^{0}(\lambda)
= f_{1}(\lambda)a(\lambda)^{2n} + f_{2}(\lambda) b(\lambda)^{2n}
\label{pseudo1}
\eeq
where the functions $f_{1}(\lambda)$, $f_{2}(\lambda)$ are due to the boundary
(this form is derived explicitly in Appendix~B: $f_1, f_2$ are determined
explicitly by (\ref{def0})--(\ref{rel}), (\ref{a})):
\be
f_{1}(\lambda) = 2x(\lambda){\cosh\mu(\lambda+i)\
\sinh\mu(\lambda+i) \over \sinh\mu(2\lambda+i)}, 
~~ \qquad
f_{j}(\lambda)=
2x_{j}(\lambda){\cosh \mu\lambda\ \sinh \mu\lambda \over
\sinh\mu(2\lambda +i)} 
\qquad j \in \{2,3 \} 
\label{pseudo1b}
\eeq
where $x(\lambda)$ is given by 
(\ref{ansatz2})  
and 
\be 
x_{2}(\lambda) = 2\sinh\mu(\lambda
+i-{ip\over 2} +i\zeta)\ \sinh\mu(\lambda +i-{ip\over 2} -i\zeta).
\label{x}
\eeq
Note that for $\zeta \rightarrow -i \infty$ we have 
$\frac{1}{x} K \stackrel{\zeta \rightarrow -i \infty}{\rightarrow} \Id$, 
$\frac{x(\lambda)-x_{2}(\lambda)}{x(\lambda)} \rightarrow 0$.

\vskip 0.1cm

\noindent {\bf Type ($ii$)} 
Now consider solution ($ii$) (\ref{K3}). The first reference state is
$|\omega_{2}\rangle$. 
The pseudo-vacuum eigenvalue takes the form (\ref{pseudo12})
\be
\Lambda^{0}(\lambda)
= f_{1}(\lambda)a(\lambda)^{2n} + f_{3}(\lambda) b(\lambda)^{2n}
\label{pseudo12}
\eeq
where    
\be
x_{3}(\lambda) ={1\over \sinh (i \mu)}\Big (\sinh (i\mu p) \cosh
\mu (2\lambda +i) -\sinh i \mu ( p +{\pi \over 2 \mu})\cosh 2\mu
(\lambda+i)\Big ) - \cosh 2i \mu \zeta
\label{x2}
\eeq
(see Appendix B for details:
applying equations (\ref{def1})--(\ref{rel2}) on (\ref{ab})). 
The important observation here is that we were able to derive 
the pseudo-vacuum eigenvalue explicitly. 

\subsection{Duality symmetries and more reference states}

The states
 $|\omega_{2}\rangle$,  $|\omega_{3}\rangle$ 
(resp. $|\omega_{+}\rangle$, $|\omega_{1}\rangle$)
are {\it not}  exact eigenstates of the transfer matrix
when the non-trivial right boundary (\ref{K3}$i$) (resp. ($ii$)) is
{\it on}
(this can be seen as a consequence of the fact that the $K^{-}$ matrix
(\ref{K3}), (($i$), ($ii$)) breaks the 
$U_{ q}(sl(2))$,
$U_{i}(sl(2))$ part of the
observed symmetry), 
but
we could have started  our construction considering
$|\omega_{1} \rangle$
(type ($i$)), 
or
$|\omega_{3}\rangle$ 
(type ($ii$)). 

Let us briefly explain why all four reference states
give the same eigenvalue when $K^- \propto \Id$.
We introduce the involutive matrices
\be
W^{(1)} = \left(
\begin{array}{cccc}
0     &0 &0 &1\\
0 &0 &1 &0\\
0 &1 &0 &0\\
1    &0 &0 &0\\
\end{array} \right)
\,,~~W^{(2)} =
\left(
\begin{array}{cccc}
0 &1 &0 &0\\
1 &0 &0 &0\\
0 &0 &0 &1\\
0 &0 &1 &0\\
\end{array} \right)
\,, ~~W^{(3)} =
\left(
\begin{array}{cccc}
0 &0 &1 &0\\
0 &0 &0 &1\\
1 &0 &0 &0\\
0 &1 &0 &0\\
\end{array} \right)
\,,\label{trans}
\eeq
Acting   
by conjugation they satisfy
\be
W_{\tilde i}^{(l)}\ W_{\tilde j}^{(l)}\ R_{\tilde i\tilde j}(\lambda;
r^{-\delta_{l1}-\delta_{l3}+\delta_{l2}}, \hat
r^{-\delta_{l1}-\delta_{l2}+\delta_{l3}})\ W_{\tilde i}^{(l)}\
W_{\tilde j}^{(l)} &=& R_{\tilde i\tilde j}(\lambda; r,\rs),
\non
\\
W_{\tilde i}^{(l)}\ M_{\tilde i}(
r^{-\delta_{l1}-\delta_{l3}+\delta_{l2}}, \hat
r^{-\delta_{l1}-\delta_{l2}+\delta_{l3}})\ W_{\tilde i}^{(l)}\
 &=& M_{\tilde i}(r,\rs)
\label{transfo}
\eeq
Here $R_{\tilde i \tilde j}(\lambda,r,\rs)$ is the formal matrix given by the
definition of $R_{ij}(\lambda)$ in \ref{A}, 
but where $r,\rs$ are
not merely shorthand as in \ref{A}, but can be manipulated as if
separate variables.
Consequently
\be
{\cal W}^{(l)}\ t(\lambda;
r^{-\delta_{l1}-\delta_{l3}+\delta_{l2}},\hat
r^{-\delta_{l1}-\delta_{l2}+\delta_{l3}})\ {\cal W}^{(l)} =
t(\lambda;r,\rs)
\eeq
where ${\cal W}^{(l)} =W_{\tilde 1}^{(l)}
\ldots W_{\tilde n}^{(l)}$  on $V_{\tilde 1} \otimes \ldots
\otimes V_{\tilde n}$ (recall $V_{\tilde i} = {\mathbb C}^4$).
We call these actions dualities.%
\footnote{just because they are involutions.}
We stress that the transformations (\ref{transfo}) leave $a(\lambda)$ and
 $b(\lambda)$ in the $R$-matrix {\it invariant}. Furthermore,
\be
{\cal W}^{(l)}\ |\omega_{+} \rangle = |\omega_{l} \rangle,
\eeq
so  
\be
{\cal W}^{(l)}\ t(\lambda;
r^{-\delta_{l1}-\delta_{l3}+\delta_{l2}},\hat
r^{-\delta_{l1}-\delta_{l2}+ \delta_{l3}})\ |\omega_{+} \rangle =
t(\lambda;r,\rs)\ |\omega_{l}\rangle,
\label{comeg}
\eeq
but it is easy to show (taking into account Appendix~B relations (\ref{a}),
 (\ref{rel})), that the \pseudovacuum\ eigenvalue
(\ref{pseudo1}) remains invariant under the `functional duality'
symmetry 
$ p \to p^{-1}$ for all $p\in \{ r,\rs,q \}$. 
This together with (\ref{comeg}) implies that the four distinct states are
degenerate   
for $K^{-} \propto \Id$. 

Only  $W^{(1)}$ `commutes' with the 
$K$-matrices (\ref{K3}$i$), (\ref{K3b}$ii$): 
\be
   W_{\tilde i}^{(1)}\ K_{\tilde i}(\lambda; -\mu)\ W_{\tilde i}^{(1)}\
    &=& K_{\tilde i}(\lambda;\mu)
\eeq
Consequently equation (\ref{comeg}) is
   valid only for $l=1$ and for the `duality' $q\to q^{-1}$ in general.
We conclude that the degeneracy is reduced by the non-trivial boundary. 
Again this corresponds to the $K$-matrices (\ref{K3}) breaking the 
observed symmetry of the model from
$U_{q}(sl(2)) \otimes U_{i}(sl(2))$ to $U_{i}(sl(2))$ (resp. $U_{q}(sl(2))$).


Let us briefly comment on the action of the 
transformations (\ref{trans}) on the
$U_{q}(sl(2))$, $U_{i}(sl(2))$  symmetries for $K^{-} \propto \Id$. 
Let $\gamma \in \{\pi,\ \sigma\}$ (\ref{gen1}),
(\ref{gen3}). 
Then by direct calculation
\be
&&{\cal W}^{(1)}\ (\gamma^{\otimes n}
(\Delta^{ n}({\cal K})))\
{\cal W}^{(1)}=\gamma^{\otimes n}( \Delta^{' n}({\cal K})), \non
\\
&&{\cal W}^{(1)}\ (\gamma^{\otimes n} (\Delta^{ n}({\cal E})))\
{\cal W}^{(1)} = \gamma^{\otimes n} (\Delta^{'n}({\cal F})), \non
\\
&&{\cal W}^{(1)}\ (\gamma^{\otimes n} (\Delta^{ n}({\cal F})))\
{\cal W}^{(1)} = \gamma^{\otimes n}( \Delta^{' n}({\cal E}))
\eeq
where $\Delta'$ is derived as (\ref{cop}) but with ${\cal K} \to
{\cal K}^{-1}$, which suggests that the $U_{q}(sl(2))$ $U_{\tilde
q}(sl(2))$ symmetry is preserved under the duality transformation
${\cal W}^{(1)}$, as long as the co-product structure becomes
$\Delta'$.
The transformations ${\cal W}^{(2)}$ and
${\cal W}^{(3)}$ modify the existing symmetry.
Indeed these transformations somehow interchange the
role between the $U_{q}(sl(2))$, $U_{i}(sl(2))$ symmetries.
Let $\tilde \pi$ be defined as in (\ref{gen1}) but with
$ i \to q$, then
\be
{\cal W}^{(2)}\ (\sigma^{\otimes n} (\Delta^{ n}(x)))\ {\cal W}^{(2)} &=& \tilde
\pi^{\otimes n} (\Delta^{'n}(x)),~~ \qquad
{\cal W}^{(3)}\ (\sigma^{\otimes n}( \Delta^{n}(x)))\ {\cal W}^{(3)} =
\tilde \pi^{\otimes n} (\Delta^{n}(x)). \qquad
\eeq


\subsection{Derivation of the eigenvalues and the Bethe ansatz}
\label{6.3}

Having obtained the pseudovacuum eigenvalues in 
section~\ref{ref states} we come to  the
derivation of the general eigenvalue. We make the following assumption for the
structure of the general eigenvalue,
\newcommand{\Agoth}{{\mathfrak A}}
\be
\Lambda(\lambda) =
f_{1}(\lambda)a(\lambda)^{2n}\Agoth_{1}(\lambda) + f_{2}(\lambda)
b(\lambda)^{2n}\Agoth_{2}(\lambda), \label{eigen}
\eeq
where
$\Agoth_{1}(\lambda)$, $\Agoth_{2}(\lambda)$ are to be determined
explicitly via the analytical Bethe ansatz method
\cite{VichirkoReshetikhin83,Reshetikhin83,Reshetikhin87,Doikou00a,Doikou00b}.

In what follows we consider
the analytical rather than  the algebraic
Bethe ansatz formulation \cite{\FT} to derive the eigenvalues
and the Bethe ansatz equations 
(i.e., basically, we observe certain constraints on the transfer
   matrix eigenvalues, which fix the form of the $\Agoth_{i}$'s). 
We first derive the Bethe ansatz equations for trivial boundary conditions
$K^{\pm}\propto\Id$, then generalize to the boundary $K^{-}$ given by
   (\ref{K3}).\\
\\
$\bullet$ {\bf $K^{\pm} \propto \Id$:}  The eigenvalues of the transfer
   matrix are analytic, i.e. $\Agoth_{i}$'s  must have common poles. 
This restriction entails certain relations between  $\Agoth_{i}$'s. 
In particular, from crossing  (\ref{cross}) 
\be
   \Agoth_{1}(-\lambda -i) =\Agoth_{2}(\lambda) \label{1}
\eeq
from the fusion relation  
(see \ref{C}) 
\be
   \Agoth_{1}(\lambda +i)\ \Agoth_{2}(\lambda) =1 
\label{2}
\eeq
and from the analyticity:
the pole $\lambda =-{ i\over 2}$  in the eigenvalue expression
(\ref{eigen})  must disappear, i.e.
\be
\Agoth_{1}(-{i\over 2})=\Agoth_{2}(-{i\over 2}). \label{3}
\eeq
From equations (\ref{1}) and (\ref{2}) we have
\be
\Agoth_{1}(\lambda)\ \Agoth_{1}(-\lambda) =1. \label{4}
\eeq
Since the $R$-matrix involves trigonometric functions only,
it is implied that $\Agoth_{i}$'s can be written as products of trigonometric
functions in the following way
\be
\Agoth_{1}(\lambda) =
\prod_{i=1}^{M_{1}} {\sinh \mu (\lambda -x_{i}^{1}) \over  \sinh
\mu (\lambda -y_{i}^{1})},
\qquad ~~
\Agoth_{2}(\lambda) =
\prod_{i=1}^{M_{2}} {\sinh \mu (\lambda -x_{i}^{2}) \over  \sinh
\mu (\lambda -y_{i}^{2})}.
\eeq
Our task now is to determine the
numbers $x_{i}^{1,2}$, $y_{i}^{1,2}$ by solving the aforementioned
constraints. By solving these  constraints we derive the form of
$\Agoth_{i}$'s ($M_{1} =M_{2}=M$ immediately from crossing,
and the fact that $\Agoth_{i}$'s must have common poles):
\be
   && \Agoth_{1}(\lambda) = \prod_{i=1}^{M} {\sinh \mu (\lambda
   -\lambda_{i} -{i\over 2}) \over  \sinh \mu (\lambda
   -\lambda_{i}+{i\over 2})}\ {\sinh \mu (\lambda
   +\lambda_{i}-{i\over 2}) \over  \sinh \mu (\lambda
   +\lambda_{i}+{i\over 2})},
   \non\\ &&
\Agoth_{2}(\lambda) =\Agoth_{1}(-\lambda-i)
\label{eigen2} 
\eeq  
Here $M$ is still an unknown integer, and $\lambda_1, \lambda_2,...$ are
   constants to be determined.
   In principle, $M$ can be determined by
   computing the asymptotic behavior of the transfer matrix for
   $t(\lambda \to \pm \infty)$. 
However, in our case this is rather an intriguing task 
because, as already mentioned, the $R$-matrix does
   not reduce to upper (lower) triangular as usual. 
Nevertheless, it is still possible to determine $M$, as we show shortly.

Let us first derive the Bethe ansatz equations which follow from the analyticity
requirements for the eigenvalues (\ref{eigen}) 
(the poles must cancel):
\be
   \Big ({\sinh \mu (\lambda_{i} +{i \over 2}) \over
   \sinh \mu (\lambda_{i} -{i \over 2})}\Big )^{2n} = \prod_{i\neq
   j=1}^{M}{\sinh \mu (\lambda_{i} -\lambda_{j} +i) \over \sinh \mu
   (\lambda_{i} -\lambda_{j} -i )}\ {\sinh \mu (\lambda_{i}
   +\lambda_{j} +i) \over \sinh \mu (\lambda_{i} +\lambda_{j} -i )}
\label{BAE}
\eeq 
Note that Bethe ansatz equations are essentially the conditions that fix the
values of $\lambda_{i}$ in expressions (\ref{eigen}), (\ref{eigen2}), provided
that $M$ is determined.
Having derived the form of the eigenvalues and the Bethe ansatz equations
we are in the position to determine $M$. 
We use Proposition~\ref{moochoo}, 
i.e. the fact that the open XXZ Hamiltonian and the open
\gemini\ Hamiltonian for $K^{\pm}\propto\Id$ have the same spectrum
(up to multiplicities) (see also \cite{MartinSaleur94a}). 
Recall \cite{Sklyanin88} that the spectrum and Bethe ansatz equations of
the  open XXZ model $K^{\pm} =\Id$ 
are given by expressions similar to (\ref{eigen}),
(\ref{eigen2}) and (\ref{BAE}) respectively, but with 
$M_{XXZ} \in \{0, \ldots ,{N\over 2} \}$. 
Now using the fact that the two Hamiltonians have the same spectrum
we can take the derivative of (\ref{eigen}) then derive the corresponding energies
along the lines in ({\ref{H0}) and compare them. 
From the comparison it immediately
follows that $M =M_{XXZ} \in  \{0, \ldots ,{N\over 2} \}$. 
Thus for the open twin spin chain with $K^{\pm} \propto \Id$, $M$ is determined. 
\\
\\
$\bullet$ {\bf $K^{-} \not\propto \Id$:}  
When $K^{-}$ is given by (\ref{K3}), (we restrict
here for type ($i$) and ($ii$) solutions) 
the eigenvalues are modified (the functions
$f_{1}$ and $f_{2}$ are changed) and
therefore the Bethe ansatz equations are modified accordingly,
namely the following extra factors:
\\
{\bf Type (i)}
\be
{\sinh \mu (\lambda_{i} +i \zeta +{ip-i \over 2})
   \over 
 \sinh \mu (\lambda_{i} +i\zeta-{ip-i \over 2})}\ 
{\sinh \mu (\lambda_{i} -i \zeta +{ip-i \over 2}) 
   \over 
 \sinh \mu (\lambda_{i} +i\zeta-{ip-i \over 2})}
\eeq 
\\
{\bf Type (ii)}
   \be
   {\sinh (i\mu p) \cosh 2\mu \lambda  -\sinh i \mu (p
   +{\pi \over 2 \mu})\cosh \mu (2\lambda-i) - \sinh  i\mu \cosh 2i
   \mu \zeta  \over  \sinh (i\mu p )\cosh 2\mu \lambda  -\sinh i \mu
   ( p +{\pi \over 2 \mu})\cosh \mu (2\lambda+i) - \sinh i\mu \cosh
   2i \mu \zeta }
   \eeq
    multiply the LHS of
   the Bethe ansatz equation (\ref{BAE}) for solutions (i) and
   (ii) respectively.

Note that in case the right boundary is non-trivial,
as in solutions (\ref{K3}) ($i$) ($ii$), 
we are restricted to the eigenvalues entailed from the subset of \pseudovacua\ 
$\{ |\omega_{+}\rangle , |\omega_{1}\rangle \}$,  
or $ \{ |\omega_{2}\rangle , |\omega_{3}\rangle \}$  
respectively (since the
other `reference' states are no longer exact eigenstates).
Of course the total number of eigenvalues is unchanged,
but the non-trivial boundary breaks the symmetry of the 
trivial boundary model, reducing the initial degeneracy. 
Thus, in addition to the eigenvalues
computed starting from $|\omega_{+}\rangle$,
($|\omega_{1}\rangle$) or $|\omega_{2}\rangle$,
($|\omega_{3}\rangle$) as \pseudovacua, there may exist further
eigenvalues  
entailed from more complicated reference eigenstates. 
One can address the computation of these eigenvalues
following the methods described in 
\cite{Nepomechie03,CaoLin03}. 

Finally the derivation of $M$ in case  $K^{-} \neq \Id $ is a more
complicated problem than $K^{-} = \Id$. 
However, by 
Proposition~\ref{moochoo} the spectrum of the open \gemini\ spin
chain is the same as the spectrum of the open XXZ spin chain with a non--diagonal
boundary (see also \cite{MartinSaleur94a}). So again one has to compare the
Hamiltonian eigenvalues by taking the derivative of the transfer matrix
eigenvalues (\ref{H0}). 
The remaining problem is, in effect, 
to derive the eigenvalues of the XXZ chain with one
non--diagonal boundary (in the `homogeneous gradation'). 
We will treat this in a separate work. 

\section{Problems and discussion}
\subsection{Bulk case $n=2$: Eigenvalues}

There are limits to the power of proposition~\ref{moochoo}. 
For example the bulk periodic case, i.e. with transfer matrix \cite{\FT} 
\be 
t(\lambda) = tr_{0}\ T_{0}(\lambda)
\eeq
cannot be expressed directly in the algebraic form (see \cite{Martin91}). 
Thus the proposition does not force the bulk 
\gemini\ chain 
to have the same spectrum as the corresponding well known XXZ chain. 
We shall show that it does not, by determining the eigenvalues of the 
$16 \times 16$ periodic $n=2$ \gemini\ transfer matrix:

{{\scriptsize
\renewcommand{\a}{\!\! a^2 \! + \!\! b^2 \!\!\!}
\renewcommand{\b}{2ab}
\newcommand{\sab}{\!\! -(s \! + \! \frac{1}{s})ab \!\!}
\newcommand{\rab}{\!\! -(r \! + \! \frac{1}{r})ab \!\!}
\newcommand{\abab}{\!\!\! (a \! - \! qb)(a \! - \! \frac{b}{q}) \!\!\!}
\[
\left( \begin{array}{cccc|cccc|cccc|cccc}  
\a&&&&&&&&&&&&&&& \\ 
&0&&&\a&&&&&&&&&&& \\ 
&&0&&&&&&\a&&&&&&& \\ 
&&&\b&&&\sab&&&\rab&&&\a&&& \\ 
\hline 
&\a&&&0&&&&&&&&&&& \\ 
&&&&&\a&&&&&&&&&& \\ 
&&&\sab&&&\b&&&\abab&&&\rab&&& \\ 
&&&&&&&0&&&&&&\a&& \\ 
\hline
&&\a&&&&&&0&&&&&&& \\ 
&&&\rab&&&\abab&&&\b&&&\sab&&& \\ 
&&&&&&&&&&\a&&&&& \\ 
&&&&&&&&&&&0&&&\a& \\ 
\hline
&&&\a&&&\rab&&&\sab&&&\b&&& \\ 
&&&&&&&\a&&&&&&0&& \\   
&&&&&&&&&&&\a&&&0& \\    
&&&&&&&&&&&&&&&\a \\ 
\end{array}\right) 
\]
}}
where 
\be 
a=a(\lambda) = \sinh \mu (\lambda +i), 
~~ \qquad
b=b(\lambda) = \sinh \mu \lambda .  
\eeq 
As before there are eigenstates of the form $|\omega_{+} \rangle  $,
and also of the form 
\be
|\omega_{5,6} \rangle 
&=& | 00 \rangle \otimes | 01 \rangle \pm | 01 \rangle \otimes | 00 \rangle 
\label{eigenstates} 
\eeq 
The corresponding eigenvalues are 
   $a^{2}(\lambda) +b^{2}(\lambda)$ (8--fold degenerate), and
$-(a^{2}(\lambda) +b^{2}(\lambda))$, (4-fold degenerate). 
The last four eigenstates are derived by diagonalizing 
the $4 \times 4$ block in $t$.
Consider possible eigenstates of the form
\be 
|\omega_{j} \rangle  \equiv   \left (\begin{array}{c}

                        1 \\

                        x  \\
                        y\\
                        z  \\

                         \end{array} \right)\,. 
\eeq 
Here $x,y,z$ must satisfy the following equations
\be 
\Big ((r+r^{-1})(x-yz) +(\hat r+\hat r^{-1})(y-xz)\Big
   )a(\lambda)b(\lambda) = \Big (a^{2}(\lambda)+b^{2}(\lambda)\Big )(1-z^{2})
\label{eq1} 
\eeq
\be
 \Big ((r+r^{-1})(1-y^{2}) +(\hat r+\hat r^{-1})(z-xy) 
  +(q+q^{-1})x\Big )a(\lambda)b(\lambda)
 = \Big (a^{2}(\lambda)+b^{2}(\lambda)\Big )(x-zy)  
\label{eq2} 
\eeq
\be
   \Big ((r+r^{-1})(z-yx) +(\hat r+\hat r^{-1})(1-x^{2}) +(q+q^{-1})y \Big
   )a(\lambda)b(\lambda)
= \Big (a^{2}(\lambda)+b^{2}(\lambda) \Big)(y-zx)  
\label{eq3}
\eeq
This yields the following eigenvalues:
For $x=y$, $z=1$ we have:
\be 
\epsilon^{+}_{1}(\lambda) = 4a(\lambda) b(\lambda) + \Big (a(\lambda)
-qb(\lambda)\Big )\Big (a(\lambda)-q^{-1}b(\lambda)\Big ),
~~~\qquad
\epsilon^{+}_{2}(\lambda) =a^{2}(\lambda)+b^{2}(\lambda)
\label{eig1} 
\eeq 
For $x=-y$, $z=-1$:
\be 
\epsilon^{-}_{1}(\lambda) = 4a(\lambda) b(\lambda) - \Big (a(\lambda)
-qb(\lambda)\Big )\Big (a(\lambda)-q^{-1}b(\lambda)\Big ),
~~~\qquad
\epsilon^{-}_{2}(\lambda) = -a^{2}(\lambda)-b^{2}(\lambda). 
\label{eig2} 
\eeq  
Compare these eigenvalues 
for the \gemini\ periodic chain with those for periodic XXZ for $n=2$: 
\be 
\epsilon_{}(\lambda) =a^{2}(\lambda)+b^{2}(\lambda)
~\mbox{(2-fold degenerate)}, 
~
\epsilon_{\pm}(\lambda) 
= 2a(\lambda) b(\lambda) \pm \Big (a(\lambda)
-qb(\lambda)\Big )\Big (a(\lambda)-q^{-1}b(\lambda) \! \Big) 
\label{ei2} 
\eeq
Ignoring the irrelevant multiplicities, it is interesting to note that 
the spectra do not coincide.  
There is often a
close relation between the bulk and open transfer matrix eigenvalues: 
the open transfer matrix eigenvalues are `doubled' compared to the bulk ones, i.e.
if the bulk eigenvalues have the form
\be 
\Lambda_{bulk}(\lambda) = a(\lambda)^{N} \Agoth_{1}(\lambda) + b(\lambda)^{N}
\Agoth_{2}(\lambda) 
\label{as1} 
\eeq
then 
\be 
\Lambda_{open}(\lambda) 
=  f_{1}(\lambda)\ a(\lambda)^{2N} \tilde \Agoth_{1}(\lambda) 
 + f_{2}(\lambda)\ b(\lambda)^{2N} \tilde \Agoth_{2}(\lambda) 
\label{as2}
\eeq 
where the functions $f_{i}$ are due to the boundaries. 
As shown in the previous section  
the spectrum  for open \gemini\ and open XXZ spin chain are the same. 
Hence, considering (\ref{as1}), (\ref{as2}), one might have 
expected a similar statement for the bulk case as well. 


\subsection{The XXZ versus the \gemini\ $R$-matrix}

Recall that the XXZ $R$-matrix is given on ${\mathbb C}^{2}\otimes {\mathbb C}^{2}$
by
\be
   R(\lambda)
   = {\cal P}\ \left( \sinh \mu (\lambda+i)\   + \sinh \mu \lambda\ U(q)
   \right)
\label{xxz}
\eeq 
Let 
\be 
\Pi:\ {\cal X}_{1} \otimes {\cal X}_{2}\ \to {\cal X}_{2} \otimes
{\cal X}_{1}, 
\eeq 
and also define 
\be 
\Delta'(x) = \Pi \circ \Delta(x), ~~x \in
{\cal A}. 
\eeq 
Then the commutation (\ref{comre1}) can be restated in the well known form
\be
R_{12}(\lambda)\ \rho^{\otimes 2}(\Delta(x))
= \rho^{\otimes 2} (\Delta'(x))\ R_{12}(\lambda),
~~~\forall x\in     \Uqsl2  . 
\label{central}
\eeq  
where recall $\rho$ is the 2d representation of $U_{q}(sl(2))$.
   We also have the following intertwining relations
   between the representations $(\rho_{\lambda} \otimes \rho_{0} )\
   \Delta(x)$ and $(\rho_{\lambda} \otimes \rho_{0})\ \Delta'(x)$:
\be
   R_{12}(\lambda)\ (\rho_{\lambda} \otimes \rho_{0} ) \Delta(x)
   = (\rho_{\lambda} \otimes \rho_{0}) \Delta'(x)\ R_{12}(\lambda) ~~\forall x\in
{\cal A}.
\label{cop3}
\eeq
(It is in this sense that the $R$-matrix is associated 
\cite{KulishReshetikhin83,Jimbo85b} with
$U_{q}({\widehat {sl(2)}})$.)

Relations (\ref{central}), (\ref{cop3}) were first introduced  in
\cite{KulishReshetikhin83,Jimbo85b}, 
establishing the quantum group approach in obtaining
solutions of the Yang-Baxter equation (\ref{YBE}). They also play
   a crucial role in the study of the underlying symmetries in 2D
   relativistic integrable field theories \cite{BernardLeclair91},
    and they have been extensively used for computing the corresponding
   exact $S$-matrices  (see e.g. \cite{BernardLeclair91}).
    Relation (\ref{cop3}) takes a simple
   form for $\lambda \to \pm \infty$, namely
\be
(\sigma^{\pm} \otimes q^{{1\over 2}\sigma^{z}}) R_{\pm}  
=  R_{\pm} (\sigma^{\pm} \otimes q^{-{1\over 2}  \sigma^{z}}),
   ~~(q^{-{1\over 2}\sigma^{z}} \otimes \sigma^{\mp}) R_{\pm}
    = R_{\pm} (q^{{1\over 2}\sigma^{z}} \otimes \sigma^{\mp}),
\eeq
where  
$R_{\pm}= R_{12}(\pm \infty)$.
Moreover, the XXZ $R$-matrix reduces to upper (lower) triangular
matrix as $\lambda \to \pm \infty$, which makes the study of the asymptotic
behavior of the transfer matrix (\ref{transfer})
 and its symmetry relatively easy.



One can show that relations of the type
   (\ref{central}) are also valid for the \gemini\ $R$-matrix (\ref{R2})
   and the representations $\rho_{i}$, $\pi$ and $\sigma$
   defined in   
(\ref{sigma}), (\ref{gen1}), (\ref{gen3}). 
Indeed let $ h \in \{ \pi,\ \sigma,\ \rho_{i} \}$ then it can be
shown by straightforward computation that 
\be 
h^{\otimes 2}(\Delta'(x))\
R_{12}(\lambda)=
   R_{12}(\lambda)\  h^{\otimes 2}(\Delta(x)), ~~~\forall x \in {\cal G}
\label{central2} 
\eeq 
where $R$ is the \gemini\ matrix.
   In fact (\ref{central2}) is just an `un$\check c$hecked' restatement of the
   various commutations.

Straightforward generalizations of the evaluation representation
(along the lines in (\ref{action})) are $\pi_{\lambda}, ~\sigma_{\lambda}, ~
\rho_{\lambda}^{1}, ~\rho_{\lambda}^{2}$.
   The derivation of a result analogous to (\ref{cop3})
   remains an open problem
   (the obvious constructions based on the representations above
   do not work).
   This means that so far
   we have not been able to see the relation of our $R$-matrix with a
   corresponding quantum affine algebra.
It is this, together with the
   fact that the $R$-matrix does {\it not} reduce to an upper (lower)
   triangular matrix for $\lambda \to \pm \infty$, 
which makes the study of the asymptotic
   behavior of the transfer matrix (\ref{transfer}) and its symmetry
   such an intriguing task.

   Although the study of the asymptotic behavior of the $R$-matrix is
   complicated for generic values of $q$, one can exploit relations
   (\ref{central}) and study the symmetry of the open transfer matrix
   along the lines described in \cite{Doikou04a,Doikou04b}.


Martin and Saleur's original tensor space $b_n$ representation 
\cite{MartinSaleur94a} has
also been discussed recently in \cite{NicholsRittenbergdeGier05}. 




   \bigskip

   \textbf{Acknowledgements:} A.D. is  supported by the European Network ``EUCLID.
   Integrable models and applications: from
   strings to
   condensed matter", contract number HPRN-CT-2002-00325.
\appendix
\setcounter{section}{0}
\renewcommand{\thesection}{Appendix \Alph{section}}

\section{The explicit $R$-matrix}\label{A}
Firstly we write $\Theta({\cal U}_{1})$ as a $16 \times 16$ matrix
acting {\em not} on 
$V_{\lef 2} \otimes V_{\lef 1} \otimes V_{\ri 1} \otimes V_{\ri 2}$
as in (\ref{the rank 1 property}), but on 
$( V_{\ri 1} \otimes V_{\lef 1} ) \otimes ( V_{\ri 2} \otimes V_{\lef 2} )$
\[
\Theta({\cal U}_{1})= 
\left( \begin{array}{cccccccccccccccc}  
0&&&&&&&&&&&&&&& \\ 
&0&&&&&&&&&&&&&& \\ 
&&0&&&&&&&&&&&&& \\ 
&&&-i^{}&&&-r^{-1}&&&-\hat r^{}&&&1&&& \\ 
&&&&0&&&&&&&&&&& \\ 
&&&&&0&&&&&&&&&& \\ 
&&&-r^{-1}&&&-q^{-1}&&&1&&&-\hat r^{-1}&&& \\ 
&&&&&&&0&&&&&&&& \\ 
&&&&&&&&0&&&&&&& \\ 
&&&-\hat r^{}&&&1&&&-q&&&-r^{}&&& \\ 
&&&&&&&&&&0&&&&& \\ 
&&&&&&&&&&&0&&&& \\ 
&&&1&&&-\hat r^{-1}&&&-r^{}&&&i&&& \\ 
&&&&&&&&&&&&&0&& \\ 
&&&&&&&&&&&&&&0& \\ 
&&&&&&&&&&&&&&&0 \\ 
\end{array}\right) 
\]
   \[
   a 1 + b \Theta(\abU_1) =
   \left( \begin{array}{cccccccccccccccc}  
   a&&&&&&&&&&&&&&& \\ 
   &a&&&&&&&&&&&&&& \\ 
   &&a&&&&&&&&&&&&& \\ 
   &&&a-i^{}b&&&-r^{-1}b&&&-\hat r^{}b&&&b&&& \\ 
   &&&&a&&&&&&&&&&& \\ 
   &&&&&a&&&&&&&&&& \\ 
   &&&-r^{-1}b&&&a-q^{-1}b&&&b&&&-\hat r^{-1}b&&& \\ 
   &&&&&&&a&&&&&&&& \\ 
   &&&&&&&&a&&&&&&& \\ 
   &&&-\hat r^{}b&&&b&&&a-qb&&&-r^{}b&&& \\ 
   &&&&&&&&&&a&&&&& \\ 
   &&&&&&&&&&&a&&&& \\ 
   &&&b&&&-\hat r^{-1}b&&&-r^{}b&&&a+ib&&& \\ 
   &&&&&&&&&&&&&a&& \\ 
   &&&&&&&&&&&&&&a& \\ 
   &&&&&&&&&&&&&&&a \\ 
   \end{array}\right) 
   \]
\[
{\mathcal P} ( a 1 + b \Theta(\abU_1)) =
\left( \begin{array}{cccc|cccc|cccc|cccc}  
a&&&&&&&&&&&&&&& \\ 
&0&&&a&&&&&&&&&&& \\ 
&&0&&&&&&a&&&&&&& \\ 
&&&b&&&-\hat r^{-1}b&&&-r^{}b&&&a+ib&&& \\ 
\hline 
&a&&&0&&&&&&&&&&& \\ 
&&&&&a&&&&&&&&&& \\ 
&&&-\hat r^{}b&&&b&&&a-qb&&&-r^{}b&&& \\ 
&&&&&&&0&&&&&&a&& \\ 
\hline
&&a&&&&&&0&&&&&&& \\ 
&&&-r^{-1}b&&&a-q^{-1}b&&&b&&&-\hat r^{-1}b&&& \\ 
&&&&&&&&&&a&&&&& \\ 
&&&&&&&&&&&0&&&a& \\ 
\hline
&&&a-i^{}b&&&-r^{-1}b&&&-\hat r^{}b&&&b&&& \\ 
&&&&&&&a&&&&&&0&& \\   
&&&&&&&&&&&a&&&0& \\    
&&&&&&&&&&&&&&&a \\ 
\end{array}\right) 
\]

The basis is $\{ ijkl \; | \; i,j,k,l \in \{1,2 \} \}$, and we will
asign the obvious standard order. 
Thus ${\mathcal P} ijkl = klij$. 
Also 
\[
M = \mbox{diagonal}(i,q^{-1},q,-i) . 
\]

It is convenient for what follows to write the above \AC\ $R$-matrix 
(cf. (\ref{R2})) as a $4 \times 4$ matrix with  $4 \times 4$ matrix entries;
and give names to the corresponding blocks of the 
monodromy matrix $T$ (\ref{mono0}): 
\be 
R(\lambda) = \left(
\begin{array}{cccc}
A(\lambda)     &B_{1}(\lambda) &B_{2}(\lambda) &B(\lambda)\\
C_{1}(\lambda) &A_{1}(\lambda) &B_{5}(\lambda) &B_{3}(\lambda)\\
C_{2}(\lambda) &C_{5}(\lambda) &A_{2}(\lambda) &B_{4}(\lambda)\\
C(\lambda)     &C_{3}(\lambda) &C_{4}(\lambda) &D(\lambda)\\
\end{array} \right),
~~ \qquad 
T_{\tilde 0}(\lambda) = \left(
\begin{array}{cccc}
{\cal A}(\lambda)     &{\cal B}_{1}(\lambda) &{\cal B}_{2}(\lambda) &{\cal
B}(\lambda)\\
{\cal C}_{1}(\lambda) &{\cal A}_{1}(\lambda) &{\cal B}_{5}(\lambda) &{\cal
B}_{3}(\lambda)\\
{\cal C}_{2}(\lambda) &{\cal C}_{5}(\lambda) &{\cal A}_{2}(\lambda) &{\cal
B}_{4}(\lambda)\\
{\cal C}(\lambda)     &{\cal C}_{3}(\lambda) &{\cal C}_{4}(\lambda) &{\cal
D}(\lambda)\\
\end{array} \right)
\,. \label{monodromyo21} 
\eeq 

\section{Pseudo-vacuum eigenvalues}
\newcommand{\ww}{ | \omega_{+} \rangle }
\newcommand{\www}{ | \omega_{2} \rangle }

Here we derive the action of the monodromy matrices and the
transfer matrix on the pseudo-vacua $|\omega_{+} \rangle$ and  $|\omega_{2}
\rangle$.

{\bf (i) The action on  $|\omega_{+} \rangle$}: 
From the action of the $R$ matrix on the pseudo-vacuum 
$|+ \rangle$ we get: $A_{i}, C_{i}, B_{5} |+\rangle =0$, 
i.e. $|\omega_{+}\rangle$ is annihilated by
   the operators ${\cal A}_{i}$, ${\cal C}_{i}$, ${\cal B}_{5}$.
   Therefore, 
\be 
\hat T_{\tilde 0}(\lambda)  \; |\omega_{+}\rangle \; 
= \left( \begin{array}{cccc}
{\cal A}(\lambda)&{\cal B}'_{1}(\lambda)&{\cal B}'_{2}(\lambda) &{\cal B}'(\lambda)\\
0                &0                 &0                    &{\cal B}'_{3}(\lambda)\\
0                &0                 &0 &{\cal B}'_{4}(\lambda)\\
0                &0                 &0                     &{\cal D}(\lambda)\\
   \end{array} \right) |\omega_{+}\rangle
   \,. 
\label{monodromyo3} 
\eeq  
Then the pseudo-vacuum eigenvalue will be
   \be 
   \Lambda^{0}(\lambda) &=& \langle \omega_{+}|\Big (-r^{-1}\hat
   r x(\lambda;m) {\cal A}^{2}- r\rs^{-1} x(\lambda;m) {\cal
   D}^{2}-r \rs^{-1} x(\lambda;m){\cal C}{\cal B}' - r^{-1} \hat
   r^{-1} x(\lambda;\yy){\cal C}_{1}{\cal B}'_{1} \non\\ & & -r \hat
   r x(\lambda;m){\cal C}_{2}{\cal B}'_{2} -r \rs^{-1}
   w^{-}(\lambda){\cal C}_{3}{\cal B}'_{3} -r \rs^{-1}
   w^{+}(\lambda){\cal C}_{4}{\cal B}'_{4}\Big )|\omega_{+}\rangle.
   \label{a} 
   \eeq
The actions of 
${\cal A}$, ${\cal D}$, ${\cal B}_{i}$, ${\cal C}_{i}$, 
${\cal B}$, ${\cal C}$ on $\ww$ are
\be 
{\cal A}(\lambda) \ww  = \prod_{l= 1}^{n} A^{\tilde l} \ww , 
~~\qquad
{\cal D}(\lambda) \ww = \prod_{ l= 1}^{n}D^{\tilde l} \ww
\label{def0} 
\eeq 
where $ A^{\tilde l} = 1\otimes 1\ldots \otimes A(\lambda) \otimes \ldots  $ etc.,
\be
{\cal C}_{1,2}(\lambda) \ww &=& \prod_{l= 1}^{n-1}A^{\tilde
l}C_{1,2}^{\tilde n} \ww , 
~~\qquad
{\cal B}_{1,2}(\lambda) \ww = \prod_{l=
1}^{n-1}A^{\tilde l}B_{1,2}^{\tilde n} \ww
\non\\
{\cal C}_{3,4}(\lambda) \ww 
  &=& \prod_{l=2}^{n}D^{\tilde l}C_{3,4}^{\tilde 1} \ww ,
~~\qquad
{\cal B}_{3,4}(\lambda) \ww = \prod_{l=2}^{n}D^{\tilde
l}B_{3,4}^{\tilde 1} \ww
\label{a0} 
\eeq 
and
\be 
{\cal C}(\lambda) \ww &=& \left( \sum_{l= 1}^{n}D^{\tilde n}
   \ldots D^{\widetilde{l+1}}C^{\tilde l}A^{\widetilde{l-1}} \ldots
   A^{\tilde 1} + \sum_{l=1}^{n-1}D^{\tilde n} \ldots
   D^{\widetilde{l+2}}C_{4}^{\widetilde{l+1}}C_{2}^{\tilde
   l}A^{\widetilde{l-1}} \ldots A^{\tilde 1} \right. 
\non\\ 
&+&  \left. \sum_{l=
   1}^{n-1}D^{\tilde n} \ldots
   D^{\widetilde{l+2}}C_{3}^{\widetilde{l+1}}C_{1}^{\tilde
   l}A^{\widetilde{l-1}} \ldots A^{\tilde 1} \right) \ww 
\label{a1} 
\eeq
\be 
{\cal B}(\lambda) \ww &=& \left( \sum_{l= 1}^{n}D^{\tilde n}
   \ldots D^{\widetilde{l+1}}B^{\tilde l}A^{\widetilde{l-1}} \ldots
   A^{\tilde 1} + \sum_{l=1}^{n-1}D^{\tilde n} \ldots
   D^{\widetilde{l+2}}B_{4}^{\widetilde{l+1}}B_{2}^{\tilde
   l}A^{\widetilde{l-1}} \ldots A^{\tilde 1} \right. 
\non\\ 
&+& \left. \sum_{l=
   1}^{n-1}D^{\tilde n} \ldots
   D^{\widetilde{l+2}}B_{3}^{\widetilde{l+1}}B_{1}^{\tilde
   l}A^{\widetilde{l-1}} \ldots A^{\tilde 1} \right) \ww
\label{a2} 
\eeq 
The primed operators    are similar to  the
operators derived in the latter equations but with the parameters 
$r,~\hat r,~q \to r^{-1},~\hat r^{-1},~q^{-1}$.
   It is also useful to derive the local action of the following operators
   on the $|+ \rangle$ state:
\be
   A^{2}|+\rangle &=& a^{2}(\lambda)|+\rangle, 
~~\qquad  
D^{2}|+\rangle  =b^{2}(\lambda)|+\rangle, 
\non\\
   C_{1}B'_{1}|+\rangle &=&a^{2}(\lambda)|+\rangle,
   ~~\qquad
C_{2}B'_{2}|+\rangle=a^{2}(\lambda)|+\rangle, 
\non \\
CB'|+\rangle&=&  \Big (a(\lambda)+r^{-1}\rs b(\lambda) \Big)^{2}|+\rangle, 
\non\\
   C_{3}B'_{3}|+\rangle&=& r^{-2}b^{2}(\lambda)|+\rangle,
   ~~\qquad
C_{4}B'_{4}|+\rangle=\rs^{2}b^{2} (\lambda)|+\rangle. 
\label{b1} 
\eeq
(the primed operators are the usual operators given in Appendix~A
   with $r,\rs,q \to r^{-1}, \rs^{-1}, q^{-1}$). 
Taking into
   account equations (\ref{def0})--(\ref{b1}) we conclude that 
\be
   {\cal A}^{2}|\omega_{+} \rangle &=&a^{2n}|\omega_{+} \rangle,
~~\qquad
   {\cal D}^{2}|\omega_{+} \rangle=b^{2n}|\omega_{+} \rangle 
\non\\
   {\cal C}_{1} {\cal B}'_{1}|\omega_{+}  \rangle &=& {\cal
   C}_{2}{\cal B}'_{2}|\omega_{+} \rangle= a^{2n}|\omega_{+} \rangle
   \non\\ 
{\cal C}_{3}{\cal B}'_{3}|\omega_{+} \rangle &=&
   r^{-2}b^{2n}|\omega_{+}  \rangle, 
~~\qquad
{\cal C}_{4}{\cal
   B}'_{4}|\omega_{+} \rangle= \rs^{2}b^{2n}|\omega_{+} \rangle
   \non\\ 
{\cal C}{\cal B}'|\omega_{+}  \rangle &=&\Big  ( (a+r^{-1}
   \rs b)^{2} {a^{2n} -b^{2n} \over a^{2} -b^{2}} + (r^{-2}+\hat
   r^{2}){b^{2}a^{2n}  -a^{2}b^{2n} \over a^{2} -b^{2}}\Big )
   |\omega_{+}\rangle. 
\label{rel} 
\eeq

{\bf (ii) The action on  $|\omega_{2} \rangle$}: 
From the action of the $R$ matrix on the second pseudo-vacuum
   $|2 \rangle $ we get: 
$A, D, C_{2}, C_{3}, B_{1}, B_{4},  B, C |2\rangle =0$. 
Therefore,  
\be 
\hat T_{\tilde 0}(\lambda)
   |\omega_{2}\rangle = \left(
   \begin{array}{cccc}
   0                 &0   &{\cal B}'_{2}(\lambda)  &0 \\
   {\cal C}'_{1}(\lambda)                  &{\cal A}_{1}(\lambda)                 
   &{\cal B}'_{5}(\lambda)                  &{\cal B}'_{3}(\lambda) \\
   0                 &0                 &{\cal A}_{2}(\lambda)  &0\\
   0                 &0                 &{\cal C}'_{4}(\lambda)                     &0\\
   \end{array} \right) |\omega_{2}\rangle
   \,. 
\label{monodromyo3b} 
\eeq  
Then the pseudo--vacuum eigenvalue will be
   \be \Lambda^{0}(\lambda) &=& \langle \omega_{2}|\Big
   (q^{-1}x(\lambda;m) {\cal A}_{1}^{2}+q  x(\lambda;m) {\cal
   A}_{2}^{2}+i x(\lambda;m){\cal B}_{1}{\cal C}_{1}' +q
   x(\lambda;\yy){\cal C}_{5}{\cal B}'_{5} \non\\ & & -i
   x(\lambda;m){\cal C}_{3}{\cal B}'_{3} +q \tilde
   w^{-}(\lambda){\cal C}_{2}{\cal B}'_{2} +q \tilde
   w^{+}(\lambda){\cal B}_{4}{\cal C}'_{4}\Big )|\omega_{2}\rangle.
   \label{ab} \eeq
   The action of ${\cal A}$, ${\cal D}$, ${\cal B}_{i}$, ${\cal
   C}_{i}$, ${\cal B}$, ${\cal C}$ on $| \omega_{2} \rangle$ are
   given in below. 
\be 
{\cal A}_{i}(\lambda) \www = \prod_{l= 1}^{n}A_{i}^{\tilde l} \www, 
\label{def1} 
\eeq 
\be 
{\cal C}_{1}(\lambda) \www
&=&\prod_{l= 1}^{n-1}A_{1}^{\tilde l}C_{1}^{\tilde n} \www , 
~~\qquad
{\cal B}_{1}(\lambda) \www
=\prod_{l=1}^{n-1}A_{1}^{\tilde l}B_{1,2}^{\tilde n} \www
\non\\ 
{\cal B}_{3}(\lambda)&=&\prod_{l=1}^{n-1}A_{1}^{\tilde l}B_{3}^{\tilde n}, 
~~\qquad
{\cal C}_{3}(\lambda)=\prod_{l=1}^{n-1}A_{1}^{\tilde l}C_{3}^{\tilde n} 
\non\\
{\cal C}_{4}(\lambda)&=&\prod_{l=2}^{n}A_{2}^{\tilde l}C_{4}^{\tilde 1}, 
~~\qquad
{\cal B}_{4}(\lambda)=\prod_{l=2}^{n}A_{2}^{\tilde l}B_{4}^{\tilde 1} 
\non\\
   {\cal B}_{2}(\lambda)&=&\prod_{l=2}^{n}A_{2}^{\tilde
   l}B_{2}^{\tilde 1}, 
~~\qquad
{\cal C}_{2}(\lambda)=\prod_{l=2}^{n}A_{2}^{\tilde l}C_{2}^{\tilde 1} 
\label{a01}
\eeq
and
   \be {\cal C}_{5}(\lambda)&=& \sum_{l= 1}^{n}A_{2}^{\tilde n}
   \ldots A_{2}^{\widetilde{l+1}}C_{5}^{\tilde l}A_{1}^{\widetilde{l-1}} \ldots
   A_{1}^{\tilde 1} + \sum_{l= 1}^{n-1}A_{2}^{\tilde n} \ldots
   A_{2}^{\widetilde{l+2}}C_{2}^{\widetilde{l+1}}B_{1}^{\tilde
   l}A_{1}^{\widetilde{l-1}} \ldots A_{1}^{\tilde 1} \non\\ &+& \sum_{l=
   1}^{n-1}A_{2}^{\tilde n} \ldots
   A_{2}^{\widetilde{l+2}}B_{4}^{\widetilde{l+1}}C_{3}^{\tilde
   l}A_{1}^{\widetilde{l-1}} \ldots A_{1}^{\tilde 1} \label{a11} \eeq
   \be {\cal B}_{5}(\lambda)&=& \sum_{l= 1}^{n}A_{2}^{\tilde n}
   \ldots A_{2}^{\widetilde{l+1}}B_{5}^{\tilde l}A_{1}^{\widetilde{l-1}} \ldots
   A_{1}^{\tilde 1} + \sum_{l= 1}^{n-1}A_{2}^{\tilde n} \ldots
   A_{2}^{\widetilde{l+2}}B_{2}^{\widetilde{l+1}}C_{1}^{\tilde
   l}A_{1}^{\widetilde{l-1}} \ldots A_{1}^{\tilde 1} \non\\ &+& \sum_{l=
   1}^{n-1}A_{2}^{\tilde n} \ldots
   A_{2}^{\widetilde{l+2}}C_{4}^{\widetilde{l+1}}B_{3}^{\tilde
   l}A_{1}^{\widetilde{l-1}} \ldots A_{1}^{\tilde 1} \label{a21} \eeq
   It is also useful to derive the local action of the following operators
   on the $|\omega^{(2)} \rangle$ state:
\be
   A_{1}^{2}|\omega_{2} \rangle &=& a^{2}(\lambda)|\omega_{2} \rangle,
   ~~A_{2}^{2}|\omega_{2} \rangle =b^{2}(\lambda)|\omega_{2} \rangle, \non\\
   C_{3}B'_{3}|\omega_{2} \rangle &=&a^{2}(\lambda)|\omega_{2} \rangle,
   ~~B_{1}C'_{1}|\omega_{2} \rangle=a^{2}(\lambda)|\omega_{2} \rangle, \non \\
C_{5}B'_{5}|\omega_{2} \rangle&=&  \Big (a(\lambda)+r^{-1} \rs^{-1} b(\lambda)\Big
   )^{2}|\omega_{2} \rangle, 
\non\\
   C_{2}B'_{2}|\omega_{2} \rangle&=& r^{-2}b^{2}(\lambda)|\omega_{2}
   \rangle, ~~B_{4}C'_{4}|\omega_{2} \rangle= \rs^{-2}b^{2}
   (\lambda)|\omega_{2} \rangle, 
\label{b11} 
\eeq 
and consequently
\be
{\cal A}_{1}^{2}|\omega_{2} \rangle &=& a^{2n}(\lambda)|\omega_{2} \rangle, 
~~\qquad
{\cal A}_{2}^{2}|\omega_{2} \rangle =b^{2nn}(\lambda)|\omega_{2} \rangle, 
\non\\
{\cal C}_{3}{\cal B}'_{3}|\omega_{2} \rangle &=&a^{2}(\lambda)|\omega_{2} \rangle,
   ~~\qquad
{\cal B}_{1}C'_{1}|\omega_{2} \rangle=a^{2n}(\lambda)|\omega_{2} \rangle, 
\non \\
   {\cal C}_{5}{\cal B}'_{5}|\omega_{2} \rangle&=&  \Big
   ((a(\lambda)+r^{-1} \rs^{-1} b(\lambda))^2{a^{2n} -b^{2n} \over
   a^{2} -b^{2}} + (r^{-2}+\hat
r^{-2}){b^{2}a^{2n}  -a^{2}b^{2n} \over a^{2} -b^{2}}\Big )|\omega_{2}\rangle, 
\non\\
   {\cal C}_{2}{\cal B}'_{2}|\omega_{2} \rangle&=&
   r^{-2}b^{2n}(\lambda)|\omega_{2} \rangle, 
~~\qquad
{\cal B}_{4}{\cal C}'_{4}|\omega_{2} \rangle
= \rs^{-2}b^{2n} (\lambda)|\omega_{2} \rangle 
\label{rel2} 
\eeq 
(note that $r^{-2} +\rs^{-2} =0$).

\section{Fusion procedure}\label{C}
Relation (\ref{2}) may be derived using the fusion formalism 
\cite{MezincescuNepomechie92b,MezincescuNepomechie92a,Doikou00a}. 
We refer the reader to
\cite{MezincescuNepomechie92b,MezincescuNepomechie92a,Doikou00a}
for the general procedure. 
Here we note (concentrating on  $K^{+}=\Id$) that this carries over to
our open \gemini\ spin chain. 

The fused transfer matrix \cite[(4.17)]{MezincescuNepomechie92b} is written as
\be 
\tilde t(\lambda) = \zeta(2\lambda+2i)\ t(\lambda)\ t(\lambda + i)
   -\delta[T(\lambda)]\ \delta[\hat T(\lambda)]\ \Delta[K^{-}(\lambda)]\
   \Delta[K^{+}(\lambda)] 
\label{fusion} 
\eeq 
where we define the quantum determinants
\be 
\delta[T(\lambda)] &=& tr_{12}\Big \{ Q_{12}\ T_{1}(\lambda)\  T_{2}(\lambda
   +i)\Big \} \non\\ \delta[\hat T(\lambda)] &=& tr_{12}\Big \{ Q_{12}\ \hat
   T_{2}(\lambda)\  \hat T_{1}(\lambda +i)\Big \} \non\\
   \Delta[K^{-}(\lambda)] &=& tr_{12} \Big \{ Q_{21}\ V_{1}\ V_{2}\
   K_{1}^{-}(\lambda)\ R_{21}(2\lambda+i)\ K_{2}^{-}(\lambda+i)\Big
   \} \non\\ \Delta[K^{+}(\lambda)] &=& tr_{12} \Big \{ Q_{12}\
   V_{1}\ V_{2}\ M_{2}^{-1}\ R_{12}(-2\lambda-3i)\ M_{2} \Big \}
\label{qudet} 
\eeq 
(here for simplicity we dropped the tilde from the indices i.e. $\tilde i \to i$). 
The key feature is the quantity $Q_{12}$: 
\be 
Q_{12} =-{1\over 2 \cosh i\mu}\Theta({\cal U}_{1}). 
\eeq
This is a $16 \times 16$ matrix but, as in the XXZ case, it is a projector onto an one-dimensional space 
(see \ref{A} or \cite{DoikouMartin03})   After some cumbersome algebra we conclude that 
\be 
\delta
   [T(\lambda)] = \delta[\hat T(\lambda)] = \zeta(\lambda +i)^{n}, 
~~
   \Delta[K^{+}(\lambda)] = g(-2\lambda -3i), 
\eeq 
and for $K^{-} =\Id$, $\; \Delta[K^{-}(\lambda)] = g(2\lambda +i)$  where  
\be
   \zeta(\lambda) &=& \sinh\mu(\lambda +i)\ \sinh\mu(-\lambda
   +i),
~~
g(\lambda)=\sinh\mu(-\lambda +i). 
\eeq 
(NB, for $K^{-}$ non-trivial (as in (\ref{K3}$i$), (\ref{K3b}$ii$)) 
the function $g$ is apparently modified.)

   Having calculated the quantum determinants the following
   expression for the fused transfer matrix holds (case $K^{\pm} =\Id$) 
\be
   \tilde t(\lambda) = \zeta(2\lambda+2i)\ t(\lambda)\ t(\lambda + i)
   -\zeta(\lambda +i)^{2n}\ g(2\lambda +i)\ g(-2\lambda -3i)
   \label{fusion1} 
\eeq
We may then return to the general formalism 
of \cite{MezincescuNepomechie92b,MezincescuNepomechie92a,Doikou00a}
to obtain (\ref{2}). 


\section{The Hamiltonian}\label{D}
It is interesting to write the Hamiltonian (\ref{H}), (\ref{H1}),
   in terms of Pauli matrices. First for $K^{-} = \Id$, 
\be
   {\cal H} &=&  -{1\over 8}\sum_{i=1}^{n-1}\Big (\sigma_{i}^{x}\
   \sigma_{i+1}^{x}+ \sigma_{i}^{y}\ \sigma_{i+1}^{y} +\cosh i
   \mu_{r}\ \sigma_{i}^{z}\ \sigma_{i+1}^{z} -\cosh i\mu_{r} I
   -\sinh
     i\mu_{r}(\sigma_{i+1}^{z}-\sigma_{i}^{z})\Big ) \non\\ &\times & \Big
   (\sigma_{i'}^{x}\
   \sigma_{(i+1)'}^{x}+ \sigma_{i'}^{y}\
   \sigma_{(i+1)'}^{y} +\cosh i \mu_{\rs}\ \sigma_{i'}^{z}\
   \sigma_{(i+1)'}^{z} -\cosh i\mu_{\rs} I +\sinh
     i\mu_{\rs}(\sigma_{(i+1)'}^{z}-\sigma_{i'}^{z})\Big ) 
\non\\
   &-& {(n -1) \cosh i \mu \over 2}I +{1\over 4 \cosh i\mu}
   I_{n} \otimes  I_{ n'} 
\eeq 
For $K^{-}\neq \Id$  we have to take into account
   the extra boundary term in the Hamiltonian 
\be 
\delta {\cal H} = -{\sinh (i\mu) \over 4 \mu x(\lambda)} {d\over
   d\lambda} K^{-}(\lambda) \vert_{\lambda =0}. 
\label{boundary}
\eeq

Let us also write the Hamiltonian of the open
spin chain (\ref{H}), (\ref{H1}) for $K^-$ given e.g. by (\ref{K3}), 
in terms of the generators we introduced in section~\ref{Boundary stable}. Define 
\be
   S^{x} ={1\over 2}(S^{+} +i S^{-}), 
~~ S^{y} ={1\over 2}(S^{+} -i S^{-}) 
\eeq 
and similarly for $\tilde S^{x,y}$.
Then
\be 
{\cal H} &=&
   -{1\over 4}\sum_{l=1}^{n-1}(S_{\tilde l}^{x}\ S_{\widetilde
     {l+1}}^{x}+ S_{\tilde l}^{y}\ s_{\widetilde {l+1}}^{y} + 4 \cosh i\mu\
     S_{\tilde l}^{z}\ S_{\widetilde {l+1}}^{z} +\cosh i\mu\  I\!\!I^{(1)})
+{\sinh i \mu \over 2}(I\!\!I_{\widetilde {n-1}}^{(1)}\ S_{\tilde n}^{z}-S_{\tilde
   1}^{z}\ I\!\!I_{\tilde 2}^{(1)}) 
\non\\
   &-&{1\over 4}\sum_{l=1}^{n-1}(\tilde S_{\tilde l}^{x}\ \tilde
   S_{\widetilde
     {l+1}}^{x}+ \tilde S_{\tilde l}^{y}\ \tilde S_{\widetilde
{l+1}}^{y})+{i \over 2}(I\!\!I_{\widetilde{n-1}}^{(2)}\ \tilde S_{\tilde n}^{z}-
   \tilde S_{\tilde 1}^{z}\ I\!\!I_{\tilde 2}^{(2)})
   \non\\
   &-& {1\over 2}\sum_{l=1}^{n-1} \Big ( r^{-1}(e_{12})_{\tilde l}\
   (e_{43})_{\widetilde {l+1}}+r^{-1} (e_{21})_{\tilde l}\
   (e_{34})_{\widetilde {l+1}} +r(e_{34})_{\tilde l}\
   (e_{21})_{\widetilde {l+1}} +
   r(e_{43})_{\tilde l}\ (e_{12})_{\widetilde {l+1}} \Big ) 
\non\\
   &-& {1\over 2}\sum_{l=1}^{n-1} \Big ( \rs (e_{13})_{\tilde l}\
   (e_{42})_{\widetilde {l+1}} +\rs (e_{31})_{\tilde l}\
   (e_{24})_{\widetilde {l+1}} +\rs^{-1}(e_{24})_{\tilde l}\
   (e_{31})_{\widetilde {l+1}} + \rs^{-1}(e_{42})_{\tilde l}\
   (e_{13})_{\widetilde {l+1}} \Big ) \non\\ &+& {1\over 4
     \cosh i \mu}\  I_{\tilde n} - {\sinh (i\mu) \over 4 x(0)}(D_{\tilde
     1}+C_{\tilde 1}), 
\eeq  
where we define
\be 
I\!\!I^{(1)} = e_{11}+e_{44}, ~~I\!\!I^{(2)}=e_{22}+e_{33}
\eeq 
and they play the role of the `unit' whenever the indices
   (1,4) and (2,3) are involved respectively. 
    Notice
   that the first line of the Hamiltonian describes exactly the XXZ
   model with open boundaries, whereas the second line gives the open
   XX model. There are also some extra `mixing' terms in the
   following lines of the form $e_{ij} \otimes e_{\bar i \bar j}$
   ($i\neq j$, $i \neq \bar j$). 
The last two terms come 
   from the right boundary interaction. They are given by 
\be
   D_{\tilde 1} = \sinh i p\mu\  I\!\!I^{(1)}, ~~C_{\tilde 1} = \sinh
   i \mu(S_{\tilde 1}^{+} +S_{\tilde 1}^{-})-2\cosh i
   \mu\ S_{\tilde 1}^{z}. 
\eeq 
Note finally that the left
   boundary interaction is trivial $K^{+} =\Id$, which is why the
   corresponding term in the Hamiltonian is proportional to unit.

\bibliographystyle{h-physrev.bst}
\bibliography{init,new31,main,main1r1}

\end{document}